\documentclass[12pt,numbers,sort&compress]{article}

\usepackage{natbib}
\usepackage{hypernat}
\usepackage{amsmath}
\usepackage{amsthm}
\usepackage{amsfonts}          
\usepackage{amssymb}           
\usepackage[all]{xy}           
\usepackage{color}
\usepackage[title]{appendix}

\usepackage[backref,colorlinks=true,linktocpage]{hyperref}

\newlength{\xtrawidth}
\newlength{\xtraheight}
\def\Z{\mathbb{Z}}
\def\fnote#1#2{\begingroup\def\thefootnote{#1}\footnote{#2}
  \addtocounter{footnote}{-1}\endgroup}
\newcommand{\Rep}[1]{\ensuremath{\mathbf{#1}}}
\newcommand{\barRep}[1]{\ensuremath{\overline{\Rep{#1}}}}
\newcommand{\Vt}{{\ensuremath{\widetilde{V}}}}
\newcommand{\Vtprime}{{\ensuremath{\widetilde{V}'}}{}}
\newcommand{\Xt}{{\ensuremath{\widetilde{X}}}}
\newcommand{\eref}[1]{eq.~\eqref{#1}}

\newcommand{\cref}[1]{Chapter~\ref{#1}}
\def\fnote#1#2{\begingroup\def\thefootnote{#1}\footnote{#2}
  \addtocounter{footnote}{-1}\endgroup}

\def\IZ{\mathbb{Z}}

\def\IP{\mathbb{P}}

\def\cO{{\mathcal O}}

\def\cF{{\mathcal F}}

\def\cL{{\mathcal L}}

\def\tx{\Xt}
\def\dP9{d\mathbb{P}_9}
\def\C{\mathbb{C}}

\def\atv{\wedge^2 \Vt}
\def\tv{\Vt}
\DeclareMathOperator{\ad}{ad}
\DeclareMathOperator{\coker}{coker}
\def\op1{{\mathcal O}_{\IP^1}}

\def\v12{V_1 \otimes V_2}
\def\b{\beta}

\newcommand{\sseq}[3]{0 \longrightarrow #1 \longrightarrow #2 \longrightarrow #3 \longrightarrow 0}
\def\z3z3{{\IZ_3 \times \IZ_3}}

\def\dirac{\slash{\! \! \! \! D}}

\newcommand{\dual}{\ensuremath{\vee}}

\DeclareMathOperator{\Ext}{Ext}
\DeclareMathOperator{\rank}{rank}
\DeclareMathOperator{\im}{im}
\newcommand{\mathemph}[1]{\textcolor{red}{\mbox{\boldmath $#1$}}}

\newcommand{\OsheafXt}{\ensuremath{\mathcal{O}_{\Xt}}}

\begin{document}
\begin{titlepage}

  \title{
    \hfill{\normalsize  UPR 1131-T} \\
    \hfill{\normalsize  hep-th/0509051} \\[1em]
    {\LARGE Heterotic Standard Model Moduli
      \author{Volker Braun$^{1,2}$, Yang-Hui He$^{1}$, Burt A.~Ovrut$^{1}$,
        and Tony Pantev$^{2}$
        \fnote{~}{vbraun,
          yanghe, ovrut@physics.upenn.edu;
          tpantev@math.upenn.edu}\\[0.5cm]
        {\normalsize $^1$
          Department of Physics,}
        {\normalsize $^2$
          Department of Mathematics}\\
        {\normalsize University of Pennsylvania} \\
        {\normalsize Philadelphia, PA 19104--6395, USA}}
    }
    \date{}
  }

  \maketitle
  \begin{abstract}
    
    In previous papers, we introduced a heterotic standard model and
    discussed its basic properties.  The Calabi-Yau threefold has, 
    generically, three K\"ahler and three complex structure
    moduli. The observable sector of this 
    vacuum has the spectrum of the MSSM with one additional pair of 
    Higgs-Higgs conjugate fields. The hidden sector has no charged matter 
    in the strongly coupled string and only minimal matter for weak coupling.
    Additionally, the spectrum of both sectors will contain vector bundle
    moduli. The exact number of such moduli was conjectured to be
    small, but was not explicitly computed. In this paper, we rectify
    this and present a formalism for computing the number of
    vector bundle moduli. Using this formalism, the number of moduli 
    in both the observable and strongly coupled hidden sectors is explicitly 
    calculated.

  \end{abstract}
  \thispagestyle{empty}
\end{titlepage}

\tableofcontents

\section{Introduction}

There is a long history in the search for realistic compactifications
of the heterotic string, see \cite{Gross:1985rr, Sen:1985eb,
  Evans:1985vb, Breit:1985ud, Candelas:1985en, Candelas:1985hv,
  Aspinwall:1987cn, Green:1987mn, Greene:1987xh, Curio:1998vu,
  Andreas:1999ty, Donagi:1999gc, Krause:2000gp, MR1824714,
  Andreas:2000pr, Andreas:2003zb, Curio:2003ur, Curio:2004pf,
  Blumenhagen:2005ga}. But until recently, finding compactifications
yielding a viable particle spectrum had resisted all efforts. In a
series of papers~\cite{HetSM1,HetSM2,HetSM3}, we presented a
``heterotic standard model'' of particle physics. Specifically, we
presented a small class of $E_{8} \times E_{8}$ heterotic superstring
vacua whose observable sectors have the spectrum of the minimal
supersymmetric standard model (MSSM), with the exception of one
additional pair of Higgs-Higgs conjugate superfields, and {\it no
  exotic multiplets}.  Such vacua occur for both weak and strong
string coupling.

Technically, our heterotic standard vacua consist of stable,
holomorphic vector bundles, $V$, with structure group $SU(4)$ over
elliptically fibered Calabi-Yau threefolds, $X$, with a $\z3z3$
fundamental group. These bundles admit a gauge connection that, in
conjunction with a Wilson line, spontaneously breaks the observable
sector $E_{8}$ gauge group down to the $SU(3)_{C} \times SU(2)_{L}
\times U(1)_{Y}$ standard model group times an additional $U(1)_{B-L}$
symmetry. The spectrum arises as the $\z3z3$ invariant sheaf
cohomology.

The existence of elliptically fibered Calabi-Yau threefolds with
${\mathbb {Z}}_{2}$ and ${\mathbb {Z}}_{2} \times {\mathbb {Z}}_{2}$
fundamental group was first demonstrated
in~\cite{SM-bundle1,SM-bundle2,SM-bundle3} and~\cite{z2z2-1,z2z2-2}
respectively.  More recently, elliptic Calabi-Yau threefolds with
$\z3z3$ fundamental group were constructed and
classified~\cite{dP9Z3Z3}. Methods for building stable, holomorphic
vector bundles with arbitrary structure group in $E_{8}$ over simply
connected elliptic Calabi-Yau threefolds were introduced
in~\cite{DonagiPrincipal, FMW1, FMW2, FMW3} and greatly expanded in a
number of papers~\cite{SM-bundle1, SM-bundle2, SM-bundle3,
  Diaconescu:1998kg, Donagi:2004qk, Donagi:2004ia}. These
constructions were then generalized to elliptically fibered Calabi-Yau
threefolds with non-trivial fundamental group
in~\cite{SM-bundle3,Ovrut:2002jk,z2z2-1,z2z2-2}.  In order to obtain a
realistic spectrum, it was found necessary to introduce a new
method~\cite{SM-bundle1,SM-bundle2,SM-bundle3,z2z2-1,z2z2-2} for
constructing vector bundles. This consists of building the requisite
bundles by ``extension'' from simpler, lower rank bundles. This method
was used for manifolds with ${\mathbb{Z}}_{2}$ fundamental group
in~\cite{SM-bundle3,SU5-z2-1,SU5-z2-2} and in the heterotic standard
model context in~\cite{dP9Z3Z3}.  In recent
work~\cite{HetSM1,HetSM2,HetSM3,SU5-z2-1,SU5-z2-2}, it was shown how
to compute the complete low-energy spectrum of such vacua. This
requires one to evaluate the relevant sheaf cohomologies, find the
action of the finite fundamental group on these spaces and, finally,
to tensor this with the action of the Wilson line on the associated
representation. The low energy spectrum is the invariant cohomology
subspaces under the resulting group action. This was applied
in~\cite{HetSM1,HetSM2,HetSM3} to compute the exact spectrum of all
multiplets transforming {\it non-trivially} under the action of the
low energy gauge group. The accompanying natural method of
``doublet-triplet'' splitting was also discussed.

Although a similar calculation in principle, the spectrum of gauge
{\it singlet} superfields was only partially determined. In addition
to the three K\"ahler and three complex structure moduli, there are
vector bundle moduli whose number was not computed. The reason is that
the relevant cohomology space lies in a complex of intertwined long
exact sequences which makes it, in general, much harder to evaluate
than the other sheaf cohomologies. Be this as it may, vector bundle
moduli are important in the particle phenomenology of these vacua,
contributing, for example, to the mu-terms and Yukawa couplings.
Furthermore, these moduli are central to the discussion of vacuum
stability~\cite{Buchbinder:2003pi, Buchbinder:2004cs, bun-mod-1,
  bun-mod-2, bun-mod-3, bun-mod-4, He:2003tj, Curio:2001qi,
  Becker:2005nb, deCarlos:2005kh}, the cosmological
constant~\cite{Buchbinder:2004nt, Buchbinder:2004im, Becker:2004gw},
and cosmology~\cite{Buchbinder:2005jy, Borisov:2005mf, Becker:2005sg}.
Hence, it is essential that their spectrum be computed.

In this paper, we present a general formalism for evaluating the
number of gauge singlet superfields for vector bundles constructed by
extension.  We then apply this method to explicitly compute the number
of vector bundle moduli in both the observable and hidden sectors of a
heterotic standard model. Specifically, we do the following. In
Section~\ref{sec:preliminaries}, all relevant properties of both the
Calabi-Yau threefold and the holomorphic vector bundles of the
heterotic standard vacua are outlined.  For clarity, we present our
discussion and formalism in terms of the observable sector vector
bundle. The hidden sector bundle will be introduced in the final
section. The sheaf cohomologies and their relation to the low-energy
spectrum are briefly discussed and the cohomology space of vector
bundle moduli is presented. The relevant short and long exact
sequences are given in Section~\ref{sec:sequences}. The various
cohomologies in the intertwined complex of long exact sequences are
systematically calculated using two Leray spectral sequences. In
Section~\ref{sec:moduli} all this information is brought together to
compute the number of vector bundle moduli. For the heterotic standard
model vacua under consideration, the number of such moduli in the
observable sector is found to be $n_{observable}=19.$ Finally, this
formalism is applied in Section~\ref{sec:hidden} to compute the number
of vector bundle moduli in the strongly coupled hidden sector. We find
that $n_{hidden}=5$. To summarize, the moduli fields are listed in
Table~\ref{tab:moduli}.
\begin{table}[htbp]
  \centering
  \begin{tabular}{c||ccccc}
    Moduli & 
    K\"ahler & 
    \begin{tabular}{c}
      Complex\\ structure
    \end{tabular} & 
    \begin{tabular}{c}
      Vector Bundle \\ (visible $E_8$)
    \end{tabular} & 
    \begin{tabular}{c}
      Vector Bundle \\ (hidden $E_8$)
    \end{tabular}
    \\ \hline
    Number & 3 & 3 & 19 & 5
  \end{tabular}
  \caption{Moduli fields in ``A Heterotic Standard Model''}
  \label{tab:moduli}
\end{table}

\section{Preliminaries}
\label{sec:preliminaries}

In our approach, there are two fundamental ingredients needed to
construct a heterotic standard model. The first is a class of
Calabi-Yau threefolds $X$ with fundamental group $\Z_3 \times \Z_3$.
The second consists of (a moduli space of) stable, holomorphic vector
bundles $V$ over $X$ with structure group $SU(4)$ which satisfy
appropriate physical constraints. Calabi-Yau threefolds of this type
were constructed in~\cite{dP9Z3Z3}.  Similarly, in~\cite{HetSM3} the
requisite holomorphic vector bundles were discussed in detail. Here,
we simply outline the properties of $X$ and $V$ that are relevant to
this paper.

\subsection{\texorpdfstring{The Calabi-Yau Threefold $X$}{The Calabi-Yau Threefold X}}
\label{s:X}

The Calabi-Yau threefold, $X$, is constructed as follows. Begin by
considering a simply connected Calabi-Yau threefold, $\Xt$, which is
an elliptic fibration over a $d\mathbb{P}_9$ surface. It was shown
in~\cite{dP9Z3Z3} that there are special $\dP9$ surfaces which admit a
$\z3z3$ action. A suitable fiber product of two such $\dP9$ surfaces
is then a Calabi-Yau threefold with an induced fixed-point free $\Z_3
\times \Z_3$ group action. Hence, the quotient $X= \Xt / (\z3z3)$ is a
smooth Calabi-Yau threefold that is torus-fibered over a singular
$\dP9$ and has non-trivial fundamental group $\Z_3 \times \Z_3$, as
desired.

Specifically, $\Xt$ is a fiber product
\begin{equation} 
  \tx = B_1 \times_{\IP^1} B_2 
  \label{tx}
\end{equation}
of two special $\dP9$ surfaces $B_1$ and $B_2$. Thus, $\Xt$
is elliptically fibered over both surfaces with the projections
\begin{equation} 
  \pi_1 : \tx \to B_1
  \,, \quad 
  \pi_2 : \tx \to B_2
  \,. 
  \label{b6}
\end{equation}
The surfaces $B_1$ and $B_2$ are themselves elliptically fibered
over $\IP^1$ with maps
\begin{equation}
  \label{betaB} 
  \b_1 : B_1 \to \IP^1
  \,, \quad 
  \b_2 : B_2 \to \IP^1
  \,.
\end{equation}
Together, these projections yield the commutative diagram 
\begin{equation}
  \label{eq:projections}
  \vcenter{\xymatrix@!0@=12mm{
      \dim_\C=3: && & \Xt \ar[dr]^{\pi_2} \ar[dl]_{\pi_1} \\
      \dim_\C=2: && B_1 \ar[dr]_{\beta_1} & & 
      B_2 \ar[dl]^{\beta_2} \\
      \dim_\C=1: && & \IP^1
      \,.
    }}
\end{equation}

The invariant homology ring of each special $d\mathbb{P}_9$ surface is
generated by two $\z3z3$ invariant curve classes $f$ and $t$ with 
intersections
\begin{equation}
  f^{2}=0
  \,, \quad 
  f t=3t^{2} = 3
  \,.
  \label{c1}
\end{equation}
Using projections~\eqref{b6}, these can be lifted to divisor classes
\begin{equation} 
  \tau_1 = \pi_1^{-1}(t_{1})
  \,, \quad 
  \tau_2 = \pi_2^{-1}(t_{2})
  \,, \quad
  \phi = \pi_1^{-1}(f_{1}) = \pi_2^{-1}(f_{2}) 
  \label{jane}
\end{equation}
on $\Xt$ satisfying the intersection relations
\begin{equation} 
  \phi^2= \tau_1^3= \tau_2^3=0
  \,, \quad 
  \phi \tau_1 = 3 \tau_1^2
  \,, \quad 
  \phi \tau_2 = 3 \tau_2^2
  \,.
\end{equation}
These three classes generate the invariant homology ring of $\Xt$.
For example, one can show that $X$ has, generically, six geometric
moduli; three K\"ahler moduli and three complex structure moduli.

Finally, the Chern classes of $\Xt$ are found to be
\begin{equation}
  \label{chernX}
  c_1\big(T\Xt\big) = c_3\big(T\Xt\big) = 0
  \,, \quad
  c_2\big(T\Xt\big) = 12(\tau_1^2 + \tau_2^2)
  \,.
\end{equation}

\subsection{\texorpdfstring{The Observable Sector Bundle $V$}{ The Observable 
SectorBundle V}}

The observable sector bundles $V$ on $X$ are produced by constructing
stable, holomorphic vector bundles $\Vt$ with structure group 
$SU(4) \subset E_{8}$
over $\Xt$ that are equivariant under the action of $\Z_3 \times
\Z_3$. Then $V=\Vt/(\z3z3)$.  One further requires that $V$ and,
hence, $\Vt$ satisfy the appropriate physical constraints.

The vector bundles $\Vt$ are constructed using a generalization of the
method of ``bundle extensions''~\cite{SM-bundle3,z2z2-2}.  Specifically,
$\Vt$ is the extension\footnote{The attentive reader will notice that
  we exchanged $V_1$ and $V_2$ in the sequence as compared
  to~\cite{HetSM1,HetSM2,HetSM3}.  This just means that we are working
  at a slightly different point in the K\"ahler and vector bundle
  moduli space. As one can easily check, the particle spectrum is
  unchanged.}
\begin{equation}
  \label{V12} 
  \sseq{V_2}{\Vt}{V_1} 
\end{equation}
of two rank two bundles $V_1$ and $V_2$ on $\Xt$. These are of the
form
\begin{equation} V_i = \cL_i \otimes \pi_2^\ast W_i, \quad i=1,2 
  \label{d3}
\end{equation}
for some line bundles $\cL_i$ on $\Xt$ and rank 2 bundles $W_i$ on
$B_2$. The rank two bundles $W_{i}$ are themselves extensions
\begin{equation} \sseq{\cO_{B_2}(a_{i} f_{2})}{W_i}{\cO_{B_2}(b_{i} f_{2}) 
    \otimes I_{k_{i}}}, 
  \label{d1} 
\end{equation}
where $a_{i},b_{i}$ are integers and $I_{k_{i}}$ is the ideal sheaf of
some $k_{i}$-tuple of points on $B_2$.

One must specify not only the bundles $\Vt$, but their transformations
under $\z3z3$ as well. To do this, first notice that for the $\z3z3$
action on the space of extensions to be well-defined, the line bundles
$\cO_{B_{2}}(a_{i}f_{2})$, $\cO_{B_{2}}(b_{i}f_{2})$ and $\cL_{i}$
must be equivariant under the finite group action. In this case, the
space of extensions will carry a representation of $\z3z3$. An
invariant class in the extension space defines an equivariant vector
bundle extension. A rank $4$ vector bundle $\Vt$ with this property
will inherit an explicit equivariant structure from the action of
$\z3z3$ on its constituent line bundles. Having found such a $\Vt$,
one can construct $V=\Vt/(\z3z3)$ on $X$.

As discussed in~\cite{HetSM1,HetSM2,HetSM3}, 
the requirement that $V$ admit a gauge connection
which satisfies the hermitian Yang-Mills equations and leads to three
chiral families of quarks/leptons, no exotic matter and two pairs of
Higgs-Higgs conjugate fields (the minimal number) imposes strong
constraints on $\Vt$. These are the following. First, in order for the
hermitian Yang-Mills gauge connection to exist on $\Vt$ this vector
bundle must be (slope) stable. A non-trivial set of necessary
conditions for stability are
\begin{equation}
  H^{0}\Big(\Xt,\Vt\Big) = H^{0}\Big(\Xt,\Vt^\dual\Big)=0
  \,, \qquad 
  H^{0}\Big(\Xt, \Vt \otimes \Vt^\dual\Big)=1
  \,.
  \label{burt1}
\end{equation}
The remaining three physical constraints were shown in~\cite{HetSM3} 
to require that
\begin{equation}
  c_{3}\big(\Vt\big)=-54
  \,, \quad 
  h^{1}\Big(\Xt,\Vt^\dual\Big)=0
  \,, \quad 
  h^{1}\Big(\Xt,\wedge^{2}\Vt\Big)=14
  \label{burt2}
\end{equation}
respectively.

A unique (up to continuous moduli) solution for $\Vt$ that is
compatible with all of our constraints\footnote{We verified that the
  cohomology groups of $\Vt$ satisfy the constraints eq.~\eqref{burt1}
  imposed by stability. We also have not been able to find any
  destabilizing subbundle, and it appears to be stable to experts in
  the field. Stability was proven for a very similar vector bundle
  in~\cite{z2z2-2}, and we expect the same methods to work in our
  case. Of course, ultimately one has to give a detailed mathematical
  proof. This will be presented elsewhere.} was found
in~\cite{HetSM3}.  It is constructed as follows.  First consider the
rank two bundles $W_{i}$ for $i=1,2$ on $B_{2}$.  Take $W_{1}$ to be
\begin{equation}
  W_{1}=\cO_{B_2} \oplus \cO_{B_2}.
  \label{f11}
\end{equation}
Note that this is the trivial extension of~\eqref{d1} with
$a_{1}=b_{1}=k_{1}=0$. Now let $W_{2}$ be an equivariant bundle in the
space of extension of the form
\begin{equation}
  \sseq{\cO_{B_2}(-2f_{2})}{W_{2}}{\chi_2\cO_{B_2}(2f_{2}) \otimes
    I_9}
  \,,
  \label{DD}
\end{equation}
where for the ideal sheaf $I_9$ of 9 points we take a generic $\z3z3$
orbit. Second, choose the two line bundles $\cL_{i}$ for $i=1,2$ on
$\Xt$ to be
\begin{equation}
  \cL_{1}= \chi_2 \cO_{\Xt}(-\tau_1 + \tau_2)
  \label{f2}
\end{equation}
and
\begin{equation}
  \cL_{2}= \cO_{\Xt}(\tau_1 - \tau_2)
  \label{f3}
\end{equation}
respectively. Here, $\chi_{1}$ and $\chi_{2}$ are the two natural
one-dimensional representations of $\z3z3$ defined by
\begin{equation}
  {\chi}_{1}(g_{1})=\omega
  \,, \quad 
  {\chi}_{1}(g_{2})=1
  \,; \qquad
  {\chi}_{2}(g_{1})=1
  \,, \quad 
  {\chi}_{2}(g_{2})=\omega
  \,,
  \label{e1}
\end{equation}
where $g_{1,2}$ are the generators of the two $\mathbb{Z}_{3}$
factors, ${\chi}_{1,2}$ are two group characters of $\z3z3$, and
$\omega = e^{\frac{2 \pi i}{3}}$ is a third root of unity.

It follows that the two rank 2 bundles $V_{1,2}$ defined in \eref{d3}
are given by
\begin{equation}
  \label{solV}
  \begin{split}
    V_1 =&~
    \chi_2 \cO_{\Xt}(-\tau_1 + \tau_2) \oplus 
    \chi_2 \cO_{\Xt}(-\tau_1 + \tau_2)
    \\
    V_2 =&~
    \cO_{\Xt}(\tau_1 - \tau_2) \otimes \pi_2^\ast W_{2}
    \,.      
  \end{split}
\end{equation}
The observable sector bundle $\Vt$ is then an equivariant element of
the space of extensions \eref{V12}.

\subsection{Computing the Particle Spectrum}

As discussed in detail in~\cite{HetSM3}, 
the low-energy particle spectrum is given by
\begin{multline}
  \label{spec} 
  \ker(\dirac_{\Vt}) = 
  \left( H^0(\tx, \cO_{\tx}) \otimes \Rep{45} \right)^{\z3z3}  
  \oplus
  \left( H^1\big(\tx, \ad(\tv) \big) 
    \otimes \Rep{1} \right)^{\z3z3} 
  \oplus \\ \oplus  
  \left( H^1(\tx, \tv) \otimes \Rep{16} \right)^{\z3z3}  \oplus 
  \left( H^1(\tx, \tv^\dual ) \otimes \barRep{16} \right)^{\z3z3} \oplus 
  \left( H^1(\tx, \atv ) \otimes \Rep{10} \right)^{\z3z3}
  \,,
\end{multline}
where the superscript indicates the invariant subspace under the
action of $\z3z3$.  The invariant cohomology space $( H^0(\tx,
\cO_{\tx}) \otimes \Rep{45} )^{\z3z3}$ corresponds to gauge
superfields in the low-energy spectrum carrying the adjoint
representation of $SU(3)_{C} \times SU(2)_{L} \times U(1)_{Y} \times
U(1)_{B-L}$.  The matter cohomology spaces, $( H^1(\tx, \tv) \otimes
\Rep{16} )^{\z3z3}$, $( H^1(\tx, \tv^\dual ) \otimes \barRep{16}
)^{\z3z3}$ and $( H^1(\tx, \atv ) \otimes \Rep{10} )^{\z3z3}$ were all
explicitly computed in~\cite{HetSM3}, leading to three chiral families
of quarks/leptons (each family with a right-handed
neutrino~\cite{Giedt:2005vx}), no exotic superfields and two
vector-like pairs of Higgs-Higgs conjugate superfields respectively.
The remaining cohomology space in \eref{spec}, namely,
\begin{equation}
  \left( H^1\big(\tx, \ad(\tv) \big)  \otimes \Rep{1} \right)^{\z3z3}, 
  \label{burt5}
\end{equation}
corresponds to the vector bundle moduli in the low-energy spectrum,
see also~\cite{Green:1987mn, Distler:1988jj, Blumenhagen:1997vt, Curio:1998bv,
  LopesCardoso:1994is, deCarlos:2005kh, Becker:2005nb}. Since
$\ad(\tv)$ is a rank $15$ vector bundle, its cohomology is much harder
to compute than the previous cohomology spaces and, for that reason,
was not evaluated in~\cite{HetSM1,HetSM2,HetSM3}.  However, vector
bundle moduli play an essential role in mu-terms, Yukawa couplings and
in the discussion of vacuum stability and the cosmological constant.
For these reasons, and to complete the spectrum, this paper will
present a formalism for computing \eref{burt5}. We will then use this
formalism to explicitly evaluate the number of vector bundle moduli in
the heterotic standard model.

\section{The Exact Sequences}
\label{sec:sequences}

It is clear from \eref{burt5} that we must compute the cohomology
space $H^{1}\big(\tx, \ad(\tv)\big)$. First, recall that the action of
the Wilson line on the $\Rep{1}$ representation is trivial. Hence, we
only need to know the $\z3z3$-invariant part of the cohomology.
Second, note that $\ad(\tv)$ is defined to be the traceless part of
$\Vt \otimes \Vt^\dual$. But the trace part is just the trivial line
bundle, whose first cohomology group vanishes. It follows that the
vector bundle moduli are precisely
\begin{equation}
  \label{eq:tensadtrace}
  H^1\Big(\tx, \ad(\tv)\Big)^{\z3z3}
  = 
  H^1\Big(\tx, \Vt \otimes \Vt^\dual \Big)^{\z3z3}
  - 
  {\underbrace{H^1\Big(\tx, \cO_\tx \Big)}_{=0}}^{\z3z3}
  \,.
\end{equation}
Therefore, the tangent space to the moduli space is $H^1\big(\tx, \Vt
\otimes \Vt^\dual \big)^{\z3z3}$. To compute this space, one must
consider complexes of interlocking exact sequences.

\subsection{Short Exact Bundle Sequences}

Recall from \eref{V12} that the vector bundle $\Vt$ is defined by the
short exact sequence of bundles
\begin{equation}
  \label{V12new} 
  \sseq{V_2}{\Vt}{V_1}
  \,. 
\end{equation}
One can tensor this sequence on the right by the bundles
$V_{1}^\dual$, $\Vt^\dual$ and $V_{2}^\dual$ to produce three new
short exact sequences which we will refer to as $(a)$, $(b)$ and $(c)$
respectively. Now take the dual of \eref{V12new}. This gives the short
exact sequence of bundles
\begin{equation}
  \label{V12dual} 
  \sseq{V_{1}^\dual}{\Vt^\dual}{V_{2}^\dual}
  \,,
\end{equation}
which we will tensor with vector bundles $V_{2}$, $\Vt$ and $V_{1}$. Of
course, the tensor product is commutative, but we will write it as
tensoring on the left. The three resulting short exact sequences will
be referred to as $(d)$, $(e)$ and $(f)$ respectively.  The six short
exact bundle sequences constructed in this manner can be written
together as the commutative diagram of exact sequences
\begin{equation}
  \label{eq:VVdualdiag}
  \vcenter{\xymatrix@=5mm{
      && \scriptstyle (d) & \scriptstyle (e) & \scriptstyle (f)
      \\ &&
      0 \ar[d] &
      0 \ar[d] &
      0 \ar[d] &
      \\
      \scriptstyle (a)&
      0 \ar[r] &
      V_2 \otimes V_1^\dual \ar[d] \ar[r] &
      \Vt \otimes V_1^\dual \ar[d] \ar[r] &
      V_1 \otimes V_1^\dual \ar[d] \ar[r] &
      0
      \\
      \scriptstyle (b)&
      0 \ar[r] &
      V_2 \otimes \Vt^\dual \ar[d] \ar[r] &
      \Vt \otimes \Vt^\dual \ar[d] \ar[r] &
      V_1 \otimes \Vt^\dual \ar[d] \ar[r] &
      0
      \\
      \scriptstyle (c)&
      0 \ar[r] &
      V_2 \otimes V_2^\dual \ar[d] \ar[r] &
      \Vt \otimes V_2^\dual \ar[d] \ar[r] &
      V_1 \otimes V_2^\dual \ar[d] \ar[r] &
      0
      \\
      &&
      0 &
      0 &
      0 &
    }}
  \,.
\end{equation}

\subsection{Long Exact Cohomology Sequences}

Each of the six short exact bundle sequences defined above gives rise
to a long exact cohomology sequence. These can also be fit together
into the complex of intertwining sequences of the form\footnote{To
  save space, we occasionally suppress the $\Xt$ in large commutative
  diagrams.}
\begin{equation}
  \label{eq:VVdualHdiag}
  \vcenter{
    \def\objectstyle{\scriptstyle}
    \xymatrix@C=4mm@R=5mm{
      &
      \vdots \ar[d] &
      \vdots \ar[d] &
      \vdots \ar[d] &
      \vdots \ar[d] &
      \vdots \ar[d] &
      \\
      \cdots \ar[r] &
      H^{i-2}(V_1 \otimes V_2^\dual) \ar[d] \ar[r] &
      H^{i-1}(V_2 \otimes V_2^\dual) \ar[d] \ar[r] &
      H^{i-1}(\Vt \otimes V_2^\dual) \ar[d] \ar[r] &
      H^{i-1}(V_1 \otimes V_2^\dual) \ar[d] \ar[r] &
      H^{i}(V_2 \otimes V_2^\dual) \ar[d] \ar[r] & \cdots
      \\
      \cdots \ar[r] &
      H^{i-1}(V_1 \otimes V_1^\dual) \ar[d] \ar[r] &
      H^i(V_2 \otimes V_1^\dual) \ar[d] \ar[r] &
      H^i(\Vt \otimes V_1^\dual) \ar[d] \ar[r] &
      H^i(V_1 \otimes V_1^\dual) \ar[d] \ar[r] &
      H^{i+1}(V_2 \otimes V_1^\dual) \ar[d] \ar[r] & \cdots
      \\
      \cdots \ar[r] &
      H^{i-1}(V_1 \otimes \Vt^\dual) \ar[d] \ar[r] &
      H^i(V_2 \otimes \Vt^\dual) \ar[d] \ar[r] &
      H^i(\Vt \otimes \Vt^\dual) \ar[d] \ar[r] &
      H^i(V_1 \otimes \Vt^\dual) \ar[d] \ar[r] &
      H^{i+1}(V_2 \otimes \Vt^\dual) \ar[d] \ar[r] & \cdots
      \\
      \cdots \ar[r] &
      H^{i-1}(V_1 \otimes V_2^\dual) \ar[d] \ar[r] &
      H^i(V_2 \otimes V_2^\dual) \ar[d] \ar[r] &
      H^i(\Vt \otimes V_2^\dual) \ar[d] \ar[r] &
      H^i(V_1 \otimes V_2^\dual) \ar[d] \ar[r] &
      H^{i+1}(V_2 \otimes V_2^\dual) \ar[d] \ar[r] & \cdots
      \\
      \cdots \ar[r] &
      H^i(V_1 \otimes V_1^\dual) \ar[d] \ar[r] &
      H^{i+1}(V_2 \otimes V_1^\dual) \ar[d] \ar[r] &
      H^{i+1}(\Vt \otimes V_1^\dual) \ar[d] \ar[r] &
      H^{i+1}(V_1 \otimes V_1^\dual) \ar[d] \ar[r] &
      H^{i+2}(V_2 \otimes V_1^\dual) \ar[d] \ar[r] & \cdots
      \\
      &
      \vdots &
      \vdots &
      \vdots &
      \vdots &
      \vdots &
      \save "3,3"."5,5"*[F--]\frm{} \restore
    }}
  \,,
\end{equation}
where the cohomology spaces in degrees $i<0$ or $i>3$ vanish for
dimension reasons. Note that the object of interest, namely,
$H^{1}\big(\tx, \Vt \otimes \Vt^\dual\big)$, occurs in this complex.
By evaluating various other cohomology spaces in these sequences, we
will be able to explicitly compute $H^{1}\big(\tx, \Vt \otimes
\Vt^\dual\big)$.

\subsection{The ``Corner'' Cohomologies}

We begin by noting that the complex is composed of a number of $3
\times 3$ blocks, each of the form 
\begin{equation}
  \label{eq:VVdualHdiagblock}
  \vcenter{
    \xymatrix{
      *+[F--]{
        C=
        H^i\Big(V_2 \otimes V_1^\dual\Big)
      } \ar[d] \ar[r] &
      H^i\Big(\Vt \otimes V_1^\dual\Big) \ar[d] \ar[r] &
      *+[F--]{
        H^i\Big(V_1 \otimes V_1^\dual\Big)
        = A
      } \ar[d] 
      \\
      H^i\Big(V_2 \otimes \Vt^\dual\Big) \ar[d] \ar[r] &
      H^i\Big(\Vt \otimes \Vt^\dual\Big) \ar[d] \ar[r] &
      H^i\Big(V_1 \otimes \Vt^\dual\Big) \ar[d]
      \\
      *+[F--]{
        D =
        H^i\Big(V_2 \otimes V_2^\dual\Big) 
      } \ar[r] &
      H^i\Big(\Vt \otimes V_2^\dual\Big) \ar[r] &
      *+[F--]{
        H^i\Big(V_1 \otimes V_2^\dual\Big)
        = B
      }
    }}
  \,,
\end{equation}
containing exclusively degree $i$ cohomology spaces. The cohomology
spaces at the corners of each block, labeled as $A$,$B$,$C$ and $D$,
are particularly amenable to evaluation, so we begin by computing
them.

\subsubsection*{\texorpdfstring{Cohomologies $A$}{Cohomologies A}}

First consider the cohomology spaces
\begin{equation}
  A = 
  H^\ast\Big(\tx, V_{1} \otimes V_{1}^\dual\Big)
  \,.
  \label{day2}
\end{equation}
It follows from \eref{solV} that $V_{1} \otimes V_{1}^\dual$ is just
the rank $4$ trivial bundle, 
\begin{equation}
  V_{1} \otimes V_{1}^\dual={\cO_{\tx}}^{\oplus 4}
  \,.
  \label{day3}
\end{equation}
Then, its cohomology spaces are
\begin{equation}
  H^\ast\Big(\tx, V_{1} \otimes V_{1}^\dual\Big)
  =
  H^\ast(\tx, \cO_{\tx})^{\oplus 4}
  \label{day4}
\end{equation}
and, therefore, 
\begin{equation}
  \label{day5}
  \begin{aligned}
    H^{0}\big(\tx, V_{1} \otimes V_{1}^\dual\big) ~&= 4
    \,,& \quad 
    H^{1}\big(\tx, V_{1} \otimes V_{1}^\dual\big) ~&= 0
    \,,
    \\
    H^{2}\big(\tx, V_{1} \otimes V_{1}^\dual\big) ~&= 0
    \,,& \quad 
    H^{3}\big(\tx, V_{1} \otimes V_{1}^\dual\big) ~&= 4
    \,,
  \end{aligned}
\end{equation}
where we have used the simplifying notation that $\C^{\oplus 4}
\equiv 1^{\oplus 4} \equiv 4$, thought of as a $\z3z3$
representation. In fact, throughout this paper we will often denote
the trivial $n$-dimensional representation by
\begin{equation}
  1^{\oplus n} \equiv n,
  \label{last}
\end{equation}
for any positive integer $n$.

\subsubsection*{\texorpdfstring{Cohomologies $B$}{Cohomologies B}}

Next, we calculate the spaces $B$ given by
\begin{equation}
  H^\ast\Big(\tx, V_{1} \otimes V_{2}^\dual\Big).
  \label{day6}
\end{equation}
For notational simplicity, we define
\begin{equation}
  \cF=V_{1} \otimes V_{2}^\dual
  \,.
  \label{day8}
\end{equation}
These cohomology spaces are much harder to compute and will be
evaluated using several applications of the Leray spectral sequence.
The first Leray sequence is associated with integrating over the
elliptic fiber of $\pi_2:\tx\to B_2$, hence pushing the cohomology
down onto the base surface $B_{2}$. In this case, one
finds\footnote{In all the spectral sequences which we are considering
  higher differentials vanish trivially. Furthermore, there are no
  extension ambiguities for $\C$-vector spaces.}
\begin{equation}
  H^{i}\Big(\tx, \cF\Big)= 
  \bigoplus_{p,q}^{p+q=i} 
  H^{p}\Big(B_{2}, R^{q}\pi_{2\ast} \cF\Big),
  \label{day7} 
\end{equation}
where the only nonvanishing entries are for $p=0,1,2$ (since
$\dim_\C B_{2}=2$) and $q=0,1$ (since the fiber of $\tx$ is an
elliptic curve).  It follows from \eref{solV} that
\begin{equation}
  \cF=\cO_{\tx}(-2\tau_{1}+2\tau_{2})^{\oplus 2}\otimes
  \pi_{2}^\ast W_2,
  \label{day9}
\end{equation}
where we have used the fact, proven in~\cite{HetSM3}, that $W_2^\dual =
\chi_2^2 W_2$. Furthermore, we see from \eref{jane} that
\begin{equation}
  \cO_{\tx}(\tau_{i})=\pi_{i}^\ast\cO_{B_{i}}(t_{i})
  \,, \quad i=1,2
  \,.
  \label{day10}
\end{equation}
Combining this with \eref{day9} implies
\begin{equation}
  \cF=
  \left[ 
    \pi_{1}^\ast\Big( \cO_{B_{1}}(-2t_{1}) \Big) 
    \otimes 
    \pi_{2}^\ast\Big(\cO_{B_{2}}(2t_{2}) \otimes W_2\Big)
  \right]^{\oplus 2}
  \,.
  \label{day11}
\end{equation}
Then, using the projection formula and the fact that
\begin{equation}
  \big( R^{q}\pi_{2\ast} \big) \circ \pi_{1}^\ast
  =
  \beta_{2}^\ast \circ \big( R^{q}\beta_{1\ast} \big)
  \,,
  \label{j1}
\end{equation}
which follows from the commutativity of the diagram
\eref{eq:projections}, one finds
\begin{equation}
  R^{q}\pi_{2\ast}\cF=
  \left[ 
    \beta_{2}^\ast R^{q}\beta_{1\ast} \Big( \cO_{B_{1}}(-2t_{1}) \Big)
    \otimes 
    \cO_{B_{2}}(2t_{2}) \otimes W_2
  \right]^{\oplus 2}.
  \label{j2}
\end{equation}
Using this expression, we can calculate each cohomology space
$H^{p}(B_{2}, R^{q}\pi_{2\ast}\cF)$ in \eref{day7}, to which we now
proceed.

Note that the cohomologies $H^{p}(B_{2}, R^{q}\pi_{2\ast}\cF)$ fill
out the $2\times 3$ tableau\footnote{Of course, the zero-th derived
  push-down is just the ordinary push-down,
  $R^0\pi_{2\ast}=\pi_{2\ast}$.}
\begin{equation}
  \label{eq:Fleraytableau}
  \vcenter{
    \def\w{34mm}
    \def\h{8mm}
    \xymatrix@C=0mm@R=0mm{
      {\scriptstyle q=1} &
      *=<\w,\h>[F]{
        H^0\big(B_{2}, R^1\pi_{2\ast}\cF\big)
      } 
      & 
      *=<\w,\h>[F]{
        H^1\big(B_{2}, R^1\pi_{2\ast}\cF\big)
      } 
      &
      *=<\w,\h>[F]{
        H^2\big(B_{2}, R^1\pi_{2\ast}\cF\big)
      } 
      \\ 
      {\scriptstyle q=0} &
      *=<\w,\h>[F]{
        H^0\big(B_{2}, \pi_{2\ast}\cF\big)
      } 
      &
      *=<\w,\h>[F]{
        H^1\big(B_{2}, \pi_{2\ast}\cF\big)
      } 
      &
      *=<\w,\h>[F]{
        H^2\big(B_{2}, \pi_{2\ast}\cF\big)
      } 
      \\ 
      & {\scriptstyle p=0} & {\scriptstyle p=1} & {\scriptstyle p=2} 
    }}
  \,.
\end{equation}
Such tableaux are very useful in keeping track of the elements of
Leray spectral sequences. As is clear from \eref{day7}, the sum
over the diagonals yields the desired cohomology of $\cF$.  Let us
first evaluate the cohomologies with $q=0$. Since the curve
$-2t_{1}$ intersects the fiber of $B_{1}$ negatively, that is,
$-2t_{1}$ has negative degree, it follows that
\begin{equation}
  R^0 \beta_{1\ast}\Big( \cO_{B_{1}}(-2t_{1}) \Big)
  =
  \beta_{1\ast}\Big( \cO_{B_{1}}(-2t_{1}) \Big)
  =
  0
  \,.
  \label{j3}
\end{equation}
Since the push-down vanishes we immediately obtain 
\begin{equation}
  H^{p}\Big(B_{2}, \pi_{2\ast}\cF\Big)=0
  \,, \quad p=0,1,2
  \label{j4}
\end{equation}
and the Leray tableau \eref{eq:Fleraytableau} becomes
\begin{equation}
  \vcenter{
    \def\w{25mm}
    \def\h{6mm}
    \xymatrix@C=0mm@R=0mm{
      {\scriptstyle q=1} &
      *=<\w,\h>[F]{
        \scriptstyle H^0\big(B_{2}, R^1\pi_{2\ast}\cF\big)
      } 
      & 
      *=<\w,\h>[F]{
        \scriptstyle H^1\big(B_{2}, R^1\pi_{2\ast}\cF\big)
      } 
      &
      *=<\w,\h>[F]{
        \scriptstyle H^2\big(B_{2}, R^1\pi_{2\ast}\cF\big)
      } 
      \\ 
      {\scriptstyle q=0} &
      *=<\w,\h>[F]{
        0
      } 
      &
      *=<\w,\h>[F]{
        0
      } 
      &
      *=<\w,\h>[F]{
        0
      } 
      \\ 
      & {\scriptstyle p=0} & {\scriptstyle p=1} & {\scriptstyle p=2} 
    }}
  \,.
\end{equation}
One must now compute the three cohomologies in the upper row,
corresponding to $q=1$. We begin by using the fact that
\begin{equation}
  R^{1} \beta_{1\ast} \cO_{B_{1}}(-2t_{1})= 
  \cO_{ \IP^{1}}(-1)^{\oplus 6}
  \,,
  \label{j5}
\end{equation}
derived in~\cite{HetSM3}.
It follows from this and \eref{j2} that
\begin{equation}
  \begin{split}
    R^{1}\pi_{2\ast}\cF =&~ 
    \left[ 
      \beta_{2}^\ast \Big( \cO_{\IP^{1}}(-1)^{\oplus 6} \Big)
      \otimes \cO_{B_{2}} (2t_{2}) \otimes W_2
    \right]^{\oplus 2} =
    \\ =&~
    \Big(
      \cO_{B_{2}} (2t_{2}-f) \otimes W_2
    \Big)^{\oplus 12}
    \,.    
  \end{split}  
  \label{j6}
\end{equation}
Using this result, we can now compute $H^{p}\big(B_{2},
R^{1}\pi_{2\ast}\cF\big)$ by pushing down onto the base $\IP^{1}$ of
$B_{2}$ using a second Leray spectral sequence. This is given for
each $p=0,1,2$ by
\begin{equation}
  H^{p}\Big(B_{2}, R^{1}\pi_{2\ast}\cF\Big)= 
  \bigoplus_{s,t}^{s+t=p} 
  H^{s} \Big(\IP^{1}, R^{t}\beta_{2\ast}(R^{1}\pi_{2\ast}\cF)\Big),
  \label{j7}
\end{equation}
where $s=0,1$ (since $\dim_\C \IP^{1}=1$) and $t=0,1$ (since the fiber
of $B_{2}$ is one dimensional). From \eref{j6} and the projection
formula, we find that
\begin{equation}
  R^{t}\beta_{2\ast}(R^{1}\pi_{2\ast}\cF)= 
  \left[ \cO_{\IP^{1}}(-1) \otimes
    R^{t}\beta_{2\ast}\Big(\cO_{B_{2}}(2t_{2}) \otimes W_2\Big)
  \right]^{\oplus 12}
  \,.
  \label{7A}
\end{equation}
Using this expression, one can calculate the cohomology spaces
$H^{s} \big(\IP^{1}, R^{t}\beta_{2\ast}(R^{1}\pi_{2\ast}\cF)\big)$
in \eref{j7}.

First note that the cohomologies $H^{s}\big(\IP^{1},
R^{t}\beta_{2\ast}(R^{1}\pi_{2\ast}\cF)\big)$ are determined by the $2 \times 2$
Leray tableau
\begin{equation}
  \label{eq:B2leray1}
  \vcenter{
    \def\w{45mm}
    \def\h{8mm}
    \xymatrix@C=0mm@R=0mm{
      {\scriptstyle t=1} &
      *=<\w,\h>[F]{
        H^0\big( \IP^1, 
        R^1\beta_{2\ast}(R^1\pi_{2\ast}\cF)  \big)
      } 
      & 
      *=<\w,\h>[F]{
        H^1\big( \IP^1,
        R^1\beta_{2\ast}(R^1\pi_{2\ast}\cF)\big)
      } 
      \\ 
      {\scriptstyle t=0} &
      *=<\w,\h>[F]{
        H^0\big( \IP^1, 
        \beta_{2\ast}(R^1\pi_{2\ast}\cF)  \big)
      } 
      &
      *=<\w,\h>[F]{
        H^0\big( \IP^1, 
        \beta_{2\ast}(R^1\pi_{2\ast}\cF)  \big)
      } 
      \\ 
      & {\scriptstyle s=0} & {\scriptstyle s=1}
    }}
  \,.
\end{equation}
Let us first evaluate the
cohomologies with $t=1$. Since $2t_{2}$ has positive degree, it
follows that
\begin{equation}
  R^{1}\beta_{2\ast}\Big(\cO_{B_{2}}(2t_{2}) \otimes W_2 \Big)=0
  \,.
  \label{j8}
\end{equation}
Therefore,
\begin{equation}
  H^{s}\Big(\IP^{1}, R^{1}\beta_{2\ast}(R^{1}\pi_{2\ast}\cF)\Big)=0
  \,, \quad 
  s=0,1
  \label{j9}
\end{equation}
and the Leray tableau \eref{eq:B2leray1} degenerates to 
\begin{equation}
  \label{eq:B2leray2}
  \vcenter{
    \def\w{35mm}
    \def\h{6mm}
    \xymatrix@C=0mm@R=0mm{
      {\scriptstyle t=1} &
      *=<\w,\h>[F]{
        0
      } 
      & 
      *=<\w,\h>[F]{
        0
      } 
      \\ 
      {\scriptstyle t=0} &
      *=<\w,\h>[F]{
        \scriptstyle
        H^0\big( \IP^1, 
        \beta_{2\ast}(R^1\pi_{2\ast}\cF)  \big)
      } 
      &
      *=<\w,\h>[F]{
        \scriptstyle
        H^0\big( \IP^1, 
        \beta_{2\ast}(R^1\pi_{2\ast}\cF)  \big)
      } 
      \\ 
      & {\scriptstyle s=0} & {\scriptstyle s=1}
    }}
  \,.
\end{equation}

One must now compute the two cohomologies in the lower row,
corresponding to $t=0$. It was shown in~\cite{HetSM3} that
\begin{equation}
  \beta_{2\ast}\Big(\cO_{B_{2}}(2t_{2}) \otimes W_2 \Big)=
  \cO_{\IP^{1}}(-2)^{\oplus 6} \oplus \cO_{\IP^{1}}^{\oplus 3} \oplus
  \cO_{\IP^{1}}(1)^{\oplus 3}
  \,.
  \label{j10}
\end{equation}
Then from \eref{7A} one finds that
\begin{equation}
  \beta_{2\ast} \Big( R^{1} \pi_{2\ast} \cF \Big)=
  \Big[ 
  \cO_{\IP^{1}}(-3)^{\oplus 2} \oplus \cO_{\IP^{1}}(-1) \oplus
  \cO_{\IP^{1}} 
  \Big]^{\oplus 36}.
  \label{j11}
\end{equation}
Clearly, then
\begin{equation}
  h^{0}\Big (\IP^{1}, \beta_{2\ast}(R^{1}\pi_{2\ast} \cF) \Big)=36.
  \label{j12}
\end{equation}
Using results from~\cite{HetSM3}, we can obtain the corresponding
$36$-dimensional $\z3z3$ representation, and conclude that
\begin{equation}
  H^{0} \Big( \IP^{1}, 
  \beta_{2\ast} (R^{1} \pi_{2\ast} \cF) 
  \Big)=RG^{\oplus 4}
  \,,
  \label{j13}
\end{equation}
where $RG$ stands for the nine-dimensional ``regular representation''
of $\z3z3$ given by
\begin{equation}
  \begin{split}
    RG 
    &= 
    \bigoplus_{0\leq n,m\leq 2} \chi_1^n \chi_2^m
    = \\ &=
    1 \oplus \chi_{1} \oplus \chi_{2} \oplus \chi_{1}^{2} \oplus \chi_{2}^{2}
    \oplus \chi_{1} \chi_{2} \oplus \chi_{1} \chi_{2}^{2} \oplus \chi_{1}^{2}
    \chi_{2} \oplus \chi_{1}^{2} \chi_{2}^{2}
    \,.    
  \end{split}
  \label{j14}
\end{equation}
Applying Serre duality on $\IP^{1}$, and using the fact that the
canonical bundle of $\IP^{1}$ is $\cO_{\IP^{1}}(-2)$, it follows
from \eref{j13} that
\begin{equation}
  H^{1}\Big (\IP^{1}, 
  \beta_{2\ast}(R^{1}\pi_{2\ast}\cF) \Big)=RG^{\oplus 16}
  \,.
  \label{j15}
\end{equation}
These results fill out the remaining entries in the Leray tableau
\eref{eq:B2leray1} for the push-down onto $\IP^{1}$. The complete
tableau is
\begin{equation}
  \label{eq:FP1lerayfinal}
  \vcenter{
    \def\w{15mm}
    \def\h{6mm}
    \xymatrix@C=0mm@R=0mm{
      {\scriptstyle t=1} &
      *=<\w,\h>[F]{
        0
      } 
      & 
      *=<\w,\h>[F]{
        0
      } 
      \\ 
      {\scriptstyle t=0} &
      *=<\w,\h>[F]{
        RG^{\oplus 4}
      } 
      &
      *=<\w,\h>[F]{
        RG^{\oplus 16}
      } 
      \\ 
      & {\scriptstyle s=0} & {\scriptstyle s=1}
    }}
  \,.
\end{equation}

Summing the diagonals in \eref{eq:FP1lerayfinal}, we can finally
evaluate the $q=1$ cohomologies $H^{p} ( B_{2}, R^{1}\pi_{2\ast}\cF )$
in the first Leray spectral sequence. Recall, that $p=0,1,2$ and that
$s+t=p$. Then
\begin{enumerate}
\item{$ p=0 ~\Rightarrow~ s=t=0$:}
  \begin{equation}
    H^{0} \Big( B_{2}, R^{1}\pi_{2\ast}\cF \Big)= RG^{\oplus 4}
    \,,
    \label{j16}
  \end{equation}
\item{$ p=1 ~\Rightarrow~ (s=0, t=1) \text{ or } (s=1, t=0)$:}
  \begin{equation}
    H^{1} \Big( B_{2}, R^{1}\pi_{2\ast}\cF \Big)= RG^{\oplus 16}
    \,,
    \label{j17}
  \end{equation}
\item{$ p=2 ~\Rightarrow~ s=t=1$:}
  \begin{equation}
    H^{2} \Big( B_{2}, R^{1}\pi_{2\ast}\cF \Big)= 0
    \,.
    \label{j18}
  \end{equation}
\end{enumerate}
Therefore the complete Leray tableau \eref{eq:Fleraytableau} for the
push-down from $\tx$ to $B_{2}$ is
\begin{equation}
  \label{eq:Fssfinal}
  \vcenter{
    \def\w{15mm}
    \def\h{6mm}
    \xymatrix@C=0mm@R=0mm{
      {\scriptstyle q=1} &
      *=<\w,\h>[F]{
        RG^{\oplus 4}
      } 
      & 
      *=<\w,\h>[F]{
        RG^{\oplus 16}
      } 
      &
      *=<\w,\h>[F]{
        0
      } 
      \\ 
      {\scriptstyle q=0} &
      *=<\w,\h>[F]{
        0
      } 
      &
      *=<\w,\h>[F]{
        0
      } 
      &
      *=<\w,\h>[F]{
        0
      } 
      \\ 
      & {\scriptstyle p=0} & {\scriptstyle p=1} & {\scriptstyle p=2} 
    }}
  \,.
\end{equation}
With this information one can, at last, compute the cohomologies $B$
given in \eref{day6}. To do this, use the entries in
eqns.~\eqref{eq:Fssfinal} and~\eqref{day7}, recalling that $m=p+q$.
The results are
\begin{equation}
  \label{eq:Fcohomology}
  \begin{aligned}
    H^{0}\Big(\tx,V_{1} \otimes V_{2}^\dual\Big)=&~ 0
    \,,& \quad
    H^{1}\Big(\tx,V_{1} \otimes V_{2}^\dual\Big)=&~ RG^{\oplus 4}
    \,, \\
    H^{2}\Big(\tx,V_{1} \otimes V_{2}^\dual\Big)=&~ RG^{\oplus 16}
    \,,& \quad
    H^{3}\Big(\tx,V_{1} \otimes V_{2}^\dual\Big)=&~ 0
    \,.    
  \end{aligned}
\end{equation}

\subsubsection*{\texorpdfstring{Cohomologies $C$}{Cohomologies C}}

Cohomologies $C$ can be computed directly from the cohomologies $B$ in
\eref{eq:Fcohomology}. To do this, one uses Serre duality, the fact
that, since $\tx$ is a Calabi-Yau manifold, its canonical bundle is
$\cO_{\tx}$ and the property that $RG$, given in \eref{j14}, is
self-dual. It follows that the Leray tableau for the push-down from
$\Xt$ to $B_2$ is
\begin{equation}
  \label{eq:Fdualssfinal}
  \vcenter{
    \def\w{15mm}
    \def\h{6mm}
    \xymatrix@C=0mm@R=0mm{
      {\scriptstyle q=1} &
      *=<\w,\h>[F]{
        0
      } 
      & 
      *=<\w,\h>[F]{
        0
      } 
      &
      *=<\w,\h>[F]{
        0
      } 
      \\ 
      {\scriptstyle q=0} &
      *=<\w,\h>[F]{
        0
      } 
      &
      *=<\w,\h>[F]{
        RG^{\oplus 16}
      } 
      &
      *=<\w,\h>[F]{
        RG^{\oplus 4}
      } 
      \\ 
      & {\scriptstyle p=0} & {\scriptstyle p=1} & {\scriptstyle p=2} 
    }}
  \,,
\end{equation}
and, therefore, 
\begin{equation}
  \label{eq:Fdualcohomology}
  \begin{aligned}
    H^{0}\Big(\tx,V_{2} \otimes V_{1}^\dual\Big)=&~ 0
    \,,& \quad
    H^{1}\Big(\tx,V_{2} \otimes V_{1}^\dual\Big)=&~ RG^{\oplus 16}
    \,, \\
    H^{2}\Big(\tx,V_{2} \otimes V_{1}^\dual\Big)=&~ RG^{\oplus 4}
    \,,& \quad
    H^{3}\Big(\tx,V_{2} \otimes V_{1}^\dual\Big)=&~ 0
    \,.    
  \end{aligned}
\end{equation}

\subsubsection*{\texorpdfstring{Cohomologies $D$}{Cohomologies D}}

Cohomologies $D$ are evaluated in much the same way as the $B$
cohomologies. However, the calculation is harder and rather
unenlightening. For these reasons, we will only state the results.
We find that the Leray tableau for the push-down from $\Xt$ to $B_2$
is 
\begin{equation}
  \label{eq:V2V2dualssfinal}
  \vcenter{
    \def\w{15mm}
    \def\h{6mm}
    \xymatrix@C=0mm@R=0mm{
      {\scriptstyle q=1} &
      *=<\w,\h>[F]{
        0
      } 
      & 
      *=<\w,\h>[F]{
        \rho_{33}
      } 
      &
      *=<\w,\h>[F]{
        1
      } 
      \\ 
      {\scriptstyle q=0} &
      *=<\w,\h>[F]{
        1
      } 
      &
      *=<\w,\h>[F]{
        \rho_{33} 
      } 
      &
      *=<\w,\h>[F]{
        0
      } 
      \\
      & {\scriptstyle p=0} & {\scriptstyle p=1} & {\scriptstyle p=2} 
    }}
  \,,
\end{equation}
where $\rho_{33}$ is a specific $33$-dimensional representation of
$\z3z3$ given by
\begin{equation}
  \rho_{33}=RG^{\oplus 3} \oplus \chi_{1} \oplus \chi_{2} \oplus
  \chi_{1}^{2} \oplus \chi_{2}^{2} \oplus \chi_{1}^{2}\chi_{2} \oplus
  \chi_{1}\chi_{2}^{2}
  \,.
  \label{j25}
\end{equation}
Therefore,
\begin{equation}
  \label{eq:V2V2dualcohomology}
  \begin{aligned}
    H^{0}\Big(\tx,V_{2} \otimes V_{2}^\dual\Big)=&~ 1
    \,,& \quad
    H^{1}\Big(\tx,V_{2} \otimes V_{2}^\dual\Big)=&~ \rho_{33}
    \,, \\
    H^{2}\Big(\tx,V_{2} \otimes V_{2}^\dual\Big)=&~ \rho_{33}
    \,,& \quad
    H^{3}\Big(\tx,V_{2} \otimes V_{2}^\dual\Big)=&~ 1
    \,.    
  \end{aligned}
\end{equation}

\section{The Long Exact Sequences}

We now systematically proceed to compute the remaining cohomology
spaces \eref{eq:VVdualHdiag} that will be required to evaluate
$H^{1}\big(\tx, \Vt \otimes \Vt^\dual\big)$. An important formula that
will be used over and over again in our analysis is the following.
Consider an exact sequence
\begin{equation}
  \dots 
  \longrightarrow 
  {\cal{U}} 
  \stackrel{f_{1}}{\longrightarrow} 
  {\cal{V}} 
  \longrightarrow 
  {\cal{W}} 
  \longrightarrow 
  {\cal{X}} 
  \stackrel{f_{2}}{\longrightarrow} 
  {\cal{Y}}  
  \longrightarrow 
  \dots 
  \,.
  \label{aa1}
\end{equation}
Then
\begin{equation}
  \dim_\C({\cal{W}})= 
  \dim_\C({\cal{V}}) + \dim_\C({\cal{X})} - \rank (f_{1}) -\rank (f_{2}).
  \label{aa2}
\end{equation}

\subsection{\texorpdfstring{The $H^0$ Cohomologies}{The Degree 0 Cohomologies}}
\label{sec:H0cohomology}

We first focus on the $3 \times 3$ block of $H^{0}$ cohomologies in
\eref{eq:VVdualHdiag}. Using the ``corner cohomologies'' computed in
the previous section, the block is
\begin{equation}
  \label{eq:VVdualH0diag}
  \vcenter{
    \xymatrix@C=5mm@R=6mm{
      &
      0 \ar[d] &
      0 \ar[d] &
      0 \ar[d] &
      \vdots \ar[d] &
      \\
      0 \ar[r] &
      0 \ar[d] \ar[r] &
      H^0\big(\Vt \otimes V_1^\dual\big) \ar[d] \ar[r] &
      4 \ar[d] \ar[r]^(0.6){d_2} &
      RG^{\oplus 16} \ar[d] \ar[r] & \cdots
      \\
      0 \ar[r] &
      H^0\big(V_2 \otimes \Vt^\dual\big) \ar[d] \ar[r] &
      H^0\big(\Vt \otimes \Vt^\dual\big) \ar[d] \ar[r] &
      H^0\big(V_1 \otimes \Vt^\dual\big) \ar[d] \ar[r] &
      H^1\big(V_2 \otimes \Vt^\dual\big) \ar[d] \ar[r] & \cdots
      \\
      0 \ar[r] &
      1 \ar[d]_{d_1} \ar[r] &
      H^0\big(\Vt \otimes V_2^\dual\big) \ar[d]_{d_3} \ar[r] &
      0 \ar[d] \ar[r] &
      \rho_{33} \ar[d] \ar[r] & \cdots
      \\
      \cdots \ar[r] &
      RG^{\oplus 16} \ar[d] \ar[r] &
      H^1\big(\Vt \otimes V_1^\dual\big) \ar[d] \ar[r] &
      0 \ar[d] \ar[r] &
      RG^{\oplus 4} \ar[d] \ar[r] & \cdots
      \\
      &
      \vdots &
      \vdots &
      \vdots &
      \vdots &
      \save "3,2"."2,3"."3,4"."4,3"*[F--]\frm{} \restore
    }}
  \,,
\end{equation}
where we have labeled coboundary maps $d_1$, $d_2$, and $d_3$. The bottom
horizontal exact sequence of this box is
\begin{equation}
  0 
  \longrightarrow 
  1 
  \longrightarrow 
  H^{0}\Big(\tx, \Vt \otimes V_{2}^\dual\Big) 
  \longrightarrow
  0 
  \longrightarrow 
  \rho_{33}
  \,.
  \label{aa3}
\end{equation}
Using formula \eref{aa2}, we find immediately that
\begin{equation}
  H^{0}\Big(\tx, \Vt \otimes V_{2}^\dual\Big)=1
  \,.
  \label{aa4}
\end{equation}
Similarly, the right hand vertical exact sequence is
\begin{equation}
  0 
  \longrightarrow 
  4 
  \longrightarrow 
  H^{0}\Big(\tx, V_{1} \otimes \Vt^\dual\Big) 
  \longrightarrow 
  0 
  \longrightarrow 
  0
  \,.
  \label{aa5}
\end{equation}
It then follows from \eref{aa2} that
\begin{equation}
  H^{0}\Big(\tx, V_{1} \otimes \Vt^\dual\Big)=4
  \,.
  \label{aa6}
\end{equation}
It remains to determine $H^{0}\big(\tx, \Vt \otimes V_{1}^\dual\big)$
and $H^{0}\big(\tx, V_2 \otimes \Vt^\dual\big)$ to complete the $H^0$
block, \eref{eq:VVdualH0diag}. To do that, we need to know the three
coboundary maps $d_{1}$, $d_{2}$, and $d_{3}$. First, consider the top
horizontal exact sequence
\begin{equation}
  0 
  \longrightarrow 
  0 
  \longrightarrow 
  H^{0}\Big(\tx, \Vt \otimes V_{1}^\dual\Big) 
  \longrightarrow
  4 
  \stackrel{d_{2}}{\longrightarrow} 
  RG^{\oplus 16}
  \,.
  \label{aa7}
\end{equation}
To evaluate $H^{0}\big(\tx, \Vt \otimes V_{1}^\dual\big)$, we note that 
\begin{equation}
  d_{2}:
  H^{0}\Big(\tx, \cO_{\tx}^{\oplus 4}\Big) 
  \to
  H^{1}\Big(\tx, V_{2} \otimes V_{1}^\dual\Big) 
  \label{aa8}
\end{equation}
is multiplication of constant sections by a choice of extension in
$\Ext_{\tx}^{1}\big(V_{1},V_{2}\big)$. For a generic choice of
extension, it follows that $d_{2}$ is an injective map. This then
implies that $\ker (d_{2})=0$ and, hence, that $\rank (d_{2})=4$. Using
this result, eqns.~\eqref{aa2} and~\eqref{aa7} give
\begin{equation}
  H^{0}\Big(\tx, \Vt \otimes V_{1}^\dual\Big) =0
  \,.
  \label{aa9}
\end{equation}
Next, consider the left hand vertical exact sequence
\begin{equation}
  0 
  \longrightarrow 
  0 
  \longrightarrow 
  H^{0}\Big(\tx, V_{2} \otimes \Vt^\dual\Big)
  \longrightarrow 
  1 
  \stackrel{d_{1}}{\longrightarrow} 
  RG^{\oplus 16}
  \,.
  \label{aa10}
\end{equation}
An identical proof implies that $\rank (d_{1})=1$ and, hence,
using \eref{aa2} we find
\begin{equation}
  H^{0}\Big(\tx, V_{2} \otimes \Vt^\dual\Big)=0
  \,.
  \label{aa11}
\end{equation}

The last unknown $H^{0}$ cohomology, $H^{0}\big(\tx, \Vt \otimes
\Vt^\dual\big)$, is contained in the middle vertical exact sequence
given by
\begin{equation}
  0 
  \longrightarrow 
  0 
  \longrightarrow 
  H^{0}\Big(\tx, \Vt \otimes \Vt^\dual\Big)
  \longrightarrow 
  1 
  \stackrel{d_{3}}{\longrightarrow} 
  H^{1}\Big(\tx, \Vt \otimes V_{1}^\dual\Big)
  \,.
  \label{aa12}
\end{equation}
It follows from \eref{aa2} that
\begin{equation}
  H^{0}\Big( \tx, \Vt \otimes \Vt^\dual\Big)=1-\rank(d_{3})\cdot 1
  \,.
  \label{aa13}
\end{equation}
Note that $\rank(d_{3})$ can be either $0$ or $1$. Were
$\rank(d_{3})=1$, then one would conclude from \eref{aa13} that
$H^{0}\big( \tx, \Vt \otimes \Vt^\dual\big)$ vanishes. But this is
impossible, because
\begin{equation}
  H^{0}\Big(\tx, \Vt \otimes \Vt^\dual\Big)=
  H^{0}\Big(\tx, \cO_{\tx} \Big) \oplus 
  H^{0}\Big(\tx, (\Vt \otimes\Vt^\dual)_\text{traceless}\Big)
  \,.
\end{equation}
Then, using $H^{0}\big(\tx, \cO_{\tx}\big)=1$ we see that $h^{0}( \tx,
\Vt \otimes \Vt^\dual) \geq 1$. Therefore, $\rank(d_{3})=0$ and
\eref{aa13} implies
\begin{equation}
  H^{0}\Big( \tx, \Vt \otimes \Vt^\dual\Big)=1
  \,.
  \label{aa14}
\end{equation}
In addition to completing the evaluation of the $H^{0}$ cohomologies,
\eref{aa14} is important since it proves that the vector bundle $\Vt$
indeed satisfies the third non-trivial stability condition listed in
\eref{burt1}.

\subsection{\texorpdfstring{The $H^1$ Cohomologies}{The Degree 1 Cohomologies}}

We now focus on the $3 \times 3$ block of $H^{1}$ cohomologies in
\eref{eq:VVdualHdiag}. Since it contains the space of moduli,
$H^{1}(\tx, \Vt \otimes \Vt^\dual)$, this is the final block that we
need to consider.  The $H^{1}$ cohomology block is
\begin{equation}
  \label{eq:VVdualH1diag}
  \vcenter{
    \xymatrix@C=3mm@R=5mm{
      &
      &
      \vdots \ar[d] &
      \vdots \ar[d] &
      \vdots \ar[d] &
      \vdots \ar[d] &
      \\
      &
      0 \ar[d] \ar[r] &
      1 \ar[d]_{d_1} \ar[r] &
      1 \ar[d] \ar[r] &
      0 \ar[d] \ar[r] &
      \rho_{33} \ar[d] \ar[r] & \cdots
      \\
      \cdots \ar[r] &
      4 \ar[d] \ar[r]_{d_2} &
      RG^{\oplus 16} \ar[d] \ar[r] &
      H^1\big(\Vt \otimes V_1^\dual\big) \ar[d] \ar[r] &
      0 \ar[d] \ar[r] &
      RG^{\oplus 4} \ar[d] \ar[r] & \cdots
      \\
      \cdots \ar[r] &
      4 \ar[d] \ar[r] &
      H^1\big(V_2 \otimes \Vt^\dual\big) \ar[d] \ar[r] &
      H^1\big(\Vt \otimes \Vt^\dual\big) \ar[d] \ar[r] &
      H^1\big(V_1 \otimes \Vt^\dual\big) \ar[d] \ar[r] &
      H^2\big(V_2 \otimes \Vt^\dual\big) \ar[d] \ar[r] & \cdots
      \\
      \cdots \ar[r] &
      0 \ar[d] \ar[r] &
      \rho_{33} \ar[d] \ar[r] &
      H^1\big(\Vt \otimes V_2^\dual\big) \ar[d] \ar[r] &
      RG^{\oplus 4} \ar[d] \ar[r] &
      \rho_{33} \ar[d] \ar[r] & \cdots
      \\
      \cdots \ar[r] &
      0 \ar[d] \ar[r] &
      RG^{\oplus 4} \ar[d] \ar[r] &
      H^2\big(\Vt \otimes V_1^\dual\big) \ar[d] \ar[r] &
      0 \ar[d] \ar[r] &
      0 \ar[d] \ar[r] & \cdots
      \\
      &
      \vdots &
      \vdots &
      \vdots &
      \vdots &
      \vdots &
      \save "3,4"."4,3"."5,4"."4,5"*[F--]\frm{} \restore
    }}
  \,,
\end{equation}
where we have inserted the ``corner'' cohomologies $A$, $B$, $C$ and
$D$ as well as the $H^{0}$ results derived above. We immediately note
that $H^{1}\big(\tx, V_{1} \otimes \Vt^\dual\big)$ lies in the right
hand vertical sequence
\begin{equation}
  0 
  \longrightarrow 
  0 
  \longrightarrow 
  H^{1}\Big(\tx, V_{1} \otimes \Vt^\dual\Big) 
  \longrightarrow 
  RG^{\oplus 4} 
  \longrightarrow 0
  \,.
  \label{bb1}
\end{equation}
It then follows from \eref{aa2} that
\begin{equation}
  H^{1}\Big(\tx, V_{1} \otimes \Vt^\dual\Big)= RG^{\oplus4}
  \,.
  \label{bb2}
\end{equation}
Similarly, one determines that 
\begin{equation}
  H^1\Big( \Xt, \Vt \otimes V_1^\dual\Big) = 
  RG^{\oplus 16} - \rank(d_2)\cdot 1 = 
  RG^{\oplus 16} - 4
  \,.
\end{equation}
We now proceed to evaluate the remaining elements in the $H^{1}$
block. To do this, it is essential that one knows the ranks of several
coboundary maps in the intertwined sequences. These are hard to
determine for the complete cohomology spaces. The problem simplifies,
however, if we restrict the complex of sequences to the $\z3z3$
invariant subspace of each cohomology space. Then, using the fact that
\begin{equation}
  RG^{\z3z3}=1
  \, \quad 
  \rho_{33}^{\z3z3}=3
  \,,
  \label{bb3}
\end{equation}
which follow from eqns.~\eqref{j14} and \eqref{j25} respectively, the
$H^{1}$ block and its nearby cohomologies simplify to
\begin{equation}
  \label{eq:VVdualH1diaginv}
  \vcenter{
    \xymatrix@C=5mm@R=8mm{
      &
      &
      \vdots \ar[d] &
      \vdots \ar[d] &
      \vdots \ar[d] &
      \vdots \ar[d] &
      \\
      &
      0 \ar[d] \ar[r] &
      1 \ar[d]_{d_1} \ar[r] &
      1 \ar[d]_{d_3} \ar[r] &
      0 \ar[d] \ar[r] &
      3 \ar[d]_{\delta_1} \ar[r] & \cdots
      \\
      \cdots \ar[r] &
      4 \ar[d] \ar[r]^(0.2){d_2} &
      16\ar[d] \ar[r] &
      12\ar[d] \ar[r] &
      0 \ar[d] \ar[r] &
      4 \ar[d] \ar[r] & \cdots
      \\
      \cdots \ar[r] &
      4 \ar[d] \ar[r] &
      \scriptstyle
      H^1(V_2 \otimes \Vt^\dual)^\z3z3 \ar[d] \ar[r] &
      \scriptstyle
      H^1(\Vt \otimes \Vt^\dual)^\z3z3 \ar[d] \ar[r] &
      4 \ar[d] \ar[r] &
      \scriptstyle
      H^2(V_2 \otimes \Vt^\dual)^\z3z3 \ar[d] \ar[r] & \cdots
      \\
      \cdots \ar[r] &
      0 \ar[d] \ar[r] &
      3 \ar[d]_{\delta_1} \ar[r] &
      \scriptstyle
      H^1(\Vt \otimes V_2^\dual)^\z3z3 \ar[d] \ar[r] &
      4 \ar[d] \ar[r]^-{\delta_1^\dual} &
      3 \ar[d] \ar[r] & \cdots
      \\
      \cdots \ar[r] &
      0 \ar[d] \ar[r] &
      4 \ar[d] \ar[r] &
      \scriptstyle
      H^2(\Vt \otimes V_1^\dual)^\z3z3 \ar[d] \ar[r] &
      0 \ar[d] \ar[r] &
      0 \ar[d] \ar[r] & \cdots
      \\
      &
      \vdots &
      \vdots &
      \vdots &
      \vdots &
      \vdots &
      \save "3,4"."4,3"."5,4"."4,5"*[F--]\frm{} \restore
    }}
  \,.
\end{equation}
Note that we have indicated two new coboundary maps $\delta_{1}$ and
$\delta_{1}^\dual$ in \eref{eq:VVdualH1diaginv}, as well as the maps
$d_1$, $d_2$, and $d_3$ introduced previously.

For the invariant cohomology subspaces, one can show 
\begin{equation}
  \delta_{1}=0
  \label{bb4}
\end{equation}
using the cup product in the Leray spectral sequence. We postpone the
details to Appendix~\ref{sec:leraycup}. It is exactly at this point
that we found it expedient to restrict to the invariant part of the
cohomologies. Noting that $\delta_1^\dual$ is the Serre dual of
$\delta_{1}$, it follows that
\begin{equation}
  \delta_1^\dual=0
  \label{bb5}
\end{equation}
as well. To compute the $H^{1}$ cohomologies, we must first know
$H^{2}(\tx, V_{2} \otimes \Vt^\dual)^{\z3z3}$. This lies in the vertical
sequence
\begin{equation}
  3 
  \stackrel{\delta_{1}}{\longrightarrow} 
  4 
  \longrightarrow  
  H^{2}\Big(\tx, V_{2} \otimes \Vt^\dual\Big)^{\z3z3} 
  \longrightarrow 
  3 
  \longrightarrow
  0
  \,.
  \label{bb6}
\end{equation}
Using \eref{bb4}, we immediately obtain 
\begin{equation}
  H^{2}\Big(\tx, V_{2} \otimes \Vt^\dual\Big)^{\z3z3}=7
  \,.
  \label{bb7}
\end{equation}
Serre duality then implies that
\begin{equation}
  H^{1}\Big(\tx, \Vt \otimes V_{2}^\dual\Big)^{\z3z3}=7
  \label{bb8}
\end{equation}
as well. Note that this is consistent with the lower horizontal
sequence in the $H^{1}$ block and \eref{bb5}.

Let us now consider the left hand vertical long exact sequence in the
$H^{1}$ block, which reads in part
\begin{equation}
  1 
  \stackrel{d_{1}}{\longrightarrow} 
  16 
  \longrightarrow 
  H^{1}\Big(\tx, V_{2} \otimes \Vt^\dual\Big)^{\z3z3} 
  \longrightarrow 
  3  
  \stackrel{\delta_{1}}{\longrightarrow} 
  4
  \,.
  \label{bb9}
\end{equation}
Using \eref{bb4}, the fact, previously established, that
$\rank(d_{1})=1$ and \eref{aa2}, we find that
\begin{equation}
  H^{1}\Big(\tx, V_{2} \otimes \Vt^\dual\Big)^\z3z3 =18
  \,. 
  \label{bb10}
\end{equation}
Serre duality then implies
\begin{equation}
  H^{2}\Big(\tx, \Vt \otimes V_{2}^\dual\Big)^\z3z3 =18
  \,. 
  \label{bb11}
\end{equation}
Putting this information back into the complex of sequences, we
arrive, finally, at
\begin{equation}
  \label{eq:VVdualH1diaginv2}
  \vcenter{
    \xymatrix@C=10mm@R=6mm{
      &
      &
      \vdots \ar[d] &
      \vdots \ar[d] &
      \vdots \ar[d] &
      \vdots \ar[d] &
      \\
      &
      0 \ar[d] \ar[r] &
      1 \ar[d]_{d_1} \ar[r] &
      1 \ar[d]_{d_3} \ar[r] &
      0 \ar[d] \ar[r] &
      3 \ar[d]_{\delta_1} \ar[r] & \cdots
      \\
      \cdots \ar[r] &
      4 \ar[d] \ar[r]^{d_2} &
      16\ar[d] \ar[r] &
      12\ar[d] \ar[r] &
      0 \ar[d] \ar[r] &
      4 \ar[d] \ar[r] & \cdots
      \\
      \cdots \ar[r] &
      4 \ar[d] \ar[r]^{d_4} &
      18 \ar[d] \ar[r] &
      H^1(\Vt \otimes \Vt^\dual)^\z3z3 \ar[d] \ar[r] &
      4 \ar[d] \ar[r]^-{\delta_2^\dual} &
      7 \ar[d] \ar[r] & \cdots
      \\
      \cdots \ar[r] &
      0 \ar[d] \ar[r] &
      3 \ar[d]_{\delta_1} \ar[r] &
      7 \ar[d]_{\delta_2} \ar[r] &
      4 \ar[d] \ar[r]^-{\delta_1^\dual} &
      3 \ar[d] \ar[r] & \cdots
      \\
      \cdots \ar[r] &
      0 \ar[d] \ar[r] &
      4 \ar[d] \ar[r] &
      4 \ar[d] \ar[r] &
      0 \ar[d] \ar[r] &
      0 \ar[d] \ar[r] & \cdots
      \\
      &
      \vdots &
      \vdots &
      \vdots &
      \vdots &
      \vdots &
      \save
      "3,3"."3,5"."5,3"."5,5"."3,4"."4,3"."5,4"."4,5"*[F--]\frm{}
      \restore
    }}
  \,.
\end{equation}
Note that we have introduced yet more coboundary maps: $d_{4}$,
$\delta_2$, and $\delta_2^\dual$ (the Serre dual of $\delta_2$).

\section{The Moduli}
\label{sec:moduli}

One can now solve for the tangent space to the moduli space,
$H^{1}(\tx, \Vt \otimes \Vt^\dual)^{\z3z3}$, of the observable sector.
Of course, the complex dimension of the tangent space equals the
number of moduli. To do this, consider the middle horizontal sequence
in \eref{eq:VVdualH1diaginv2} given by
\begin{equation}
  0 
  \longrightarrow 
  1 
  \longrightarrow 
  4 
  \stackrel{d_{4}}{\longrightarrow} 
  18 
  \longrightarrow 
  H^{1}\Big(\tx, \Vt \otimes \Vt^\dual\Big)^{\z3z3} 
  \longrightarrow 
  4 
  \stackrel{\delta_{2}^\dual}{\longrightarrow} 
  7
  \,.
  \label{bb12}
\end{equation}
One must now determine the rank of the coboundary maps $d_{4}$ and
$\delta_{2}^\dual$. Since we are restricted to the invariant cohomology
subspaces, one can apply methods identical to those used in
Appendix~\ref{sec:leraycup} to prove \eref{bb4}. Again, one finds that
\begin{equation}
  \delta_{2}^\dual=0
  \,.
  \label{bb13}
\end{equation}
The rank of $d_{4}$ can be determined by the exactness of the sequence
\eref{bb12}. The beginning of this sequence is
\begin{equation}
  0 
  \stackrel{\phi_{1}}{\longrightarrow} 
  1 
  \stackrel{\phi_{2}}{\longrightarrow} 
  4 
  \stackrel{d_{4}}{\longrightarrow} 
  18  
  \,,
  \label{bb14}
\end{equation}
where we named the first two maps $\phi_{1}$ and $\phi_{2}$.
Exactness implies that $\im(\phi_{1})=\ker(\phi_{2})$ and, hence, that
$\ker(\phi_{2})=0$. It follows that $\phi_{2}$ is injective and that
$\im(\phi_{2})=\ker(d_{4})=1$. Therefore, $\rank(d_{4})$ is the
difference $4-1=3$. That is,
\begin{equation}
  \rank(d_{4})=3
  \,.
  \label{bb15}
\end{equation}
Then, using eqns.~\eqref{bb13},~\eqref{bb15}, and~\eqref{aa2}, the
exact sequence \eref{bb12} tells us that the number of 
moduli of the observable sector vector bundle $V=\Vt/(\z3z3)$ is
\begin{equation}
  n_\text{observable}= h^{1}\Big(\tx, \Vt \otimes \Vt^\dual\Big)^{\z3z3}=19
  \,.
  \label{bb16}
\end{equation}

\section{\texorpdfstring{The Hidden Sector Moduli}{The Hidden Sector Moduli}}
\label{sec:hidden}

In the previous section, we computed the number of vector bundle
moduli in the observable $E_8$ gauge sector. However, there is also
the $E_8'$ hidden sector (in the following, the prime will always
denote hidden sector quantities), which potentially contributes moduli
fields to the low energy effective action. These moduli interact only
gravitationally with the fields of the standard model and, therefore,
are not immediately relevant.  Nevertheless, we would like to compute
the hidden sector moduli in this section. The reason is twofold.
First, the stability and dynamics of the hidden sector vector bundles
is important for the discussion of supersymmetry breaking via $E_{8}'$
fermion condensation and, potentially, for cosmology.  The second
reason is that the computation uses exactly the same formalism as for
the observable sector bundles. It serves, therefore, as another,
simpler, example of our method. For specificity, we will consider the
hidden sector of the strongly coupled heterotic string only.  Our
formalism is easily applied to the weak coupling case as well.

Recall that in~\cite{HetSM3}, for the case of strong string coupling,
we chose the $E_{8}'$ hidden sector gauge bundle to be an $SU(2)$
instanton $V'$ over the Calabi-Yau threefold $X$. As usual, we work
with the $\z3z3$-equivariant bundle $\Vtprime$ on the universal
covering space $\Xt$. The bundle $\Vtprime$ was explicitly defined by
the extension
\begin{equation}
  \label{eq:V12prime} 
  \sseq{V_2'}\Vtprime{V_1'} 
  \,,
\end{equation}
where $V_1'$ and $V_2'$ are the line bundles
\begin{equation}
  V_2' = \OsheafXt(2\tau_1 + \tau_2 - \phi)
  \,, \quad
  V_1' = \big(V_2'\big)^\dual = 
  \OsheafXt(-2\tau_1 - \tau_2 + \phi)
  \,.
\end{equation}

Analogous to \eref{eq:VVdualdiag}, we find that
$\Vtprime\otimes\Vtprime^\dual$ lives in a $3\times 3$ square of short exact
sequences
\begin{equation}
  \label{eq:VVprimedualdiag}
  \vcenter{\xymatrix@=3mm{
      &
      0 \ar[d] &
      0 \ar[d] &
      0 \ar[d] &
      \\
      0 \ar[r] &
      *+[F--]{
        \OsheafXt(4\tau_1 + 2\tau_2 - 2\phi) 
      }\ar[d] \ar[r] &
      \Vtprime \otimes {V_1'}^\dual \ar[d] \ar[r] &
      *+[F--]{
        \OsheafXt
      } \ar[d] \ar[r] &
      0
      \\
      0 \ar[r] &
      V_2' \otimes \Vtprime^\dual \ar[d] \ar[r] &
      \Vtprime \otimes \Vtprime^\dual \ar[d] \ar[r] &
      V_1' \otimes \Vtprime^\dual \ar[d] \ar[r] &
      0
      \\
      0 \ar[r] &
      *+[F--]{
        \OsheafXt 
      } \ar[d] \ar[r] &
      \Vtprime \otimes {V_2'}^\dual \ar[d] \ar[r] &
      *+[F--]{
        \OsheafXt(-4\tau_1 - 2\tau_2 + 2\phi)
      } \ar[d] \ar[r] &
      0
      \\
      &
      0 &
      0 &
      0
    }}
  \,.
\end{equation}
We already computed the ``corner cohomologies''
in~\cite{HetSM3}. Here, we simply quote the result that
\begin{equation}
  H^p\Big( \Xt, 
  \OsheafXt(4\tau_1 + 2\tau_2 - 2\phi) 
  \Big) = 
  H^{3-p}\Big( \Xt, 
  \OsheafXt(-4\tau_1 - 2\tau_2 + 2\phi) 
  \Big) = 
  \begin{cases}
    RG^{\oplus 6} & p=1 \\
    0 & p\not=1
    \,.
  \end{cases}
\end{equation}
Therefore, the $H^0$ cohomology block reads
\begin{equation}
  \label{eq:VVprimedualH0diag}
  \vcenter{
    \xymatrix@C=3mm@R=4mm{
      &
      0 \ar[d] &
      0 \ar[d] &
      0 \ar[d] &
      \vdots \ar[d] &
      \\
      0 \ar[r] &
      0 \ar[d] \ar[r] &
      H^0\big(\Vtprime \otimes {V_1'}^\dual\big) \ar[d] \ar[r] &
      1 \ar[d] \ar[r]^(0.6){d_2'} &
      RG^{\oplus 6} \ar[d] \ar[r] & \cdots
      \\
      0 \ar[r] &
      H^0\big(V_2' \otimes \Vtprime^\dual\big) \ar[d] \ar[r] &
      H^0\big(\Vtprime \otimes \Vtprime^\dual\big) \ar[d] \ar[r] &
      H^0\big(V_1' \otimes \Vtprime^\dual\big) \ar[d] \ar[r] &
      H^1\big(V_2' \otimes \Vtprime^\dual\big) \ar[d] \ar[r] & \cdots
      \\
      0 \ar[r] &
      1 \ar[d]_{d_1'} \ar[r] &
      H^0\big(\Vtprime \otimes {V_2'}^\dual\big) \ar[d] \ar[r] &
      0 \ar[d] \ar[r] &
      0 \ar[d] \ar[r] & \cdots
      \\
      \cdots \ar[r] &
      RG^{\oplus 6} \ar[d] \ar[r] &
      H^1\big(\Vtprime \otimes {V_1'}^\dual\big) \ar[d] \ar[r] &
      0 \ar[d] \ar[r] &
      0 \ar[d] \ar[r] & \cdots
      \\
      &
      \vdots &
      \vdots &
      \vdots &
      \vdots &
      \save "3,2"."2,3"."3,4"."4,3"*[F--]\frm{} \restore
    }}
  \,,
\end{equation}
and, using exactly the same reasoning as in
Subsection~\ref{sec:H0cohomology}, we find that $d_1'$ and $d_2'$ are
injective. Exactness of the sequence then implies that
\begin{equation}
  \begin{gathered}
    H^0\Big(\Xt, \Vtprime \otimes {V_1'}^\dual\Big)
    = 
    H^0\Big(\Xt, V_2' \otimes \Vtprime^\dual\Big)
    = 0
    \,,
    \\
    H^0\Big(\Xt, V_1' \otimes \Vtprime^\dual\Big) =
    H^0\Big(\Xt, \Vtprime \otimes {V_2'}^\dual\Big) =
    H^0\Big(\Xt, \Vtprime \otimes \Vtprime^\dual\Big) =
    1 
    \,.        
  \end{gathered}  
\end{equation}
We proceed to the $H^1$ cohomology block, which now becomes
\begin{equation}
  \label{eq:VVprimedualH1diag}
  \vcenter{
    \xymatrix@C=5mm@R=5mm{
      & &&& 
      \vdots \ar[d] &
      \vdots \ar[d] &
      \vdots \ar[d] 
      \\
      &
      0 \ar[d] & 
      0 \ar[d] & 
      0 \ar[d] \ar[r] &
      1 \ar[d]_(0.4){d_1'} \ar[r] &
      1 \ar[d] \ar[r] &
      0 \ar[d] \ar[r] &
      \cdots
      \\
      0 \ar[r] &
      0 \ar[r] \ar[d] & 0 \ar[r] \ar[d] & 1 \ar[d] \ar[r]_{d_2'} &      
      RG^{\oplus 6} \ar[d] \ar[r] &
      H^1\big(\Vtprime \otimes {V_1'}^\dual\big) \ar[d] \ar[r] &
      0 \ar[d] \ar[r] &
      \cdots
      \\
      0 \ar[r] &
      0 \ar[r] \ar[d] & 1 \ar[r] \ar[d] & 1 \ar[r] \ar[d] &
      H^1\big(V_2' \otimes \Vtprime^\dual\big) \ar[d] \ar[r] &
      H^1\big(\Vtprime \otimes \Vtprime^\dual\big) \ar[d] \ar[r] &
      H^1\big(V_1' \otimes \Vtprime^\dual\big) \ar[d] \ar[r] &
      \cdots
      \\
      0 \ar[r] &
      1 \ar[r] \ar[d] & 1 \ar[r] \ar[d] & 0 \ar[r] \ar[d] &
      0 \ar[d] \ar[r] &
      H^1\big(\Vtprime \otimes {V_2'}^\dual\big) \ar[d] \ar[r] &
      0 \ar[d] \ar[r] &
      \cdots
      \\
      &
      \vdots &
      \vdots &
      \vdots &
      \vdots &
      \vdots &
      \vdots 
      \save "4,2"."4,4"."3,3"."5,3"*[F--]\frm{} \restore
      \save "4,5"."4,7"."3,6"."5,6"*[F--]\frm{} \restore
    }}
  \,,
\end{equation}
Since we already determined that $d_1'$ and $d_2'$ inject, and
therefore
\begin{equation}
  \coker( d_1' ) = \coker( d_2' ) = RG^{\oplus{6}} - 1
  \,,
\end{equation}
we can directly read off that
\begin{equation}
  \begin{gathered}
    H^1\Big(\Xt, \Vtprime \otimes {V_1'}^\dual\Big) = 
    H^1\Big(\Xt, V_2' \otimes \Vtprime^\dual\Big) = 
    RG^{\oplus{6}} - 1
    \,,
    \\
    H^1\Big(\Xt, V_1' \otimes \Vtprime^\dual\Big) = 
    H^1\Big(\Xt, \Vtprime \otimes {V_2'}^\dual\Big) = 0    
  \end{gathered}
\end{equation}
from the long exact sequences \eref{eq:VVprimedualH1diag}. Finally,
the middle horizontal long exact sequence
\begin{equation}
  0 
  \longrightarrow
  0 
  \longrightarrow
  1
  \stackrel{\sim}{\longrightarrow}
  1 
  \stackrel{0}{\longrightarrow}
  RG^{\oplus{6}} - 1
  \stackrel{\sim}{\longrightarrow}
  H^1\big(\Xt, \Vtprime \otimes \Vtprime^\dual\big)  
  \longrightarrow
  0
  \longrightarrow
  \cdots
\end{equation}
yields
\begin{equation}
  \label{eq:Vtprimemoduli}
  H^1\Big(\Xt, \Vtprime \otimes \Vtprime^\dual\Big)   = 
  RG^{\oplus{6}} - 1
  \,.
\end{equation}
Hence, the number of vector bundle moduli of $V'=\Vtprime/(\z3z3)$ is
the invariant part of \eref{eq:Vtprimemoduli}. Using \eref{bb3}), we
find that
\begin{equation}
  \label{eq:Vtprimemoduliinv}
  n_{hidden}=h^1\Big(\Xt, \Vtprime \otimes \Vtprime^\dual\Big)^\z3z3 =
  6 - 1 = 5
  \,.
\end{equation}
We conclude that there are $5$ vector bundle moduli in the hidden
sector of the strongly coupled string.

\section*{Acknowledgments}

We are grateful to B.~Nelson for enlightening discussions. This
research was supported in part by the Department of Physics and the
Math/Physics Research Group at the University of Pennsylvania under
cooperative research agreement DE-FG02-95ER40893 with the
U.~S.~Department of Energy and an NSF Focused Research Grant
DMS0139799 for ``The Geometry of Superstrings.'' T.~P.~is partially
supported by an NSF grant DMS 0403884.

\begin{appendices}

\section{\texorpdfstring{The Coboundary Map $\delta_1$}{The Coboundary Map delta1}}
\label{sec:leraycup}

The purpose of this Appendix is to determine the coboundary map
\begin{equation}
  \delta_1:
  H^1\Big(\Xt, V_2 \otimes V_{2}^\dual\Big)^\z3z3 \to 
  H^2\Big(\Xt, V_2 \otimes V_{1}^\dual\Big)^\z3z3
\end{equation}
associated with the short exact sequence of equivariant vector bundles
(see \eref{V12dual})
\begin{equation}
  \sseq
  {V_2 \otimes V_{1}^\dual}
  {V_2 \otimes \Vt^\dual}
  {V_2 \otimes V_{2}^\dual}
  \,.
\end{equation}
The choice of extension $V_2\otimes \Vt^\dual$ is precisely the choice
of an element $x$ of the $\Ext$-space
\begin{equation}
  x \in \Ext_{\Xt}^1 \Big( V_2 \otimes V_{2}^\dual ,~ 
  V_2 \otimes V_{1}^\dual \Big)^\z3z3
  = 
  H^1\Big( \Xt, V_{2} \otimes V_{1}^\dual \Big)^\z3z3  
  \,.
\end{equation}
Therefore, $x$ must determine the coboundary map $\delta_1$. One finds
that it is the cup product, that is, the usual wedge product combined
with a suitable contraction of vector bundle indices,
\begin{equation}
  \delta_1:
  H^1\Big(\Xt, V_2 \otimes V_{2}^\dual\Big)^\z3z3 \to 
  H^2\Big(\Xt, V_2 \otimes V_{1}^\dual\Big)^\z3z3
  \,,~
  v \mapsto v \wedge x
  \,.
\end{equation}
Note that the cohomology degree is additive. Since $x$ is a degree $1$
cohomology class, the image of $\delta_1$ is indeed of degree $2$.
Nevertheless, we claim that the product map
\begin{equation}
  \label{eq:cup}
  \wedge: 
  \underbrace{
    H^1\Big(\Xt, V_2 \otimes V_2^\dual \Big)^\z3z3 
  }_{\owns~ v}
  \otimes
  \underbrace{
    H^1\Big(\Xt, V_2 \otimes V_1^\dual \Big)^\z3z3  
  }_{\owns~ x}
  \to 
  \underbrace{
  H^2\Big(\Xt, V_2 \otimes V_1^\dual \Big)^\z3z3  
  }_{\owns~ \delta_1(v) = v \wedge x}
\end{equation}
vanishes because of a refined degree stemming from the elliptic
fibration. This can be seen as follows. Let us determine the
cohomology spaces using the Leray spectral sequences
eqns.~\eqref{eq:Fdualssfinal} and \eqref{eq:V2V2dualssfinal}
corresponding to the $\pi_2:\Xt\to B_2$ fibration. First, note that
the cohomology always comes from the $\pi_{2\ast}$ push-down, and not
the $R^1\pi_{2\ast}$ part:
\begin{equation}
  \begin{aligned}
    H^1\Big(\Xt, V_2 \otimes V_{2}^\dual\Big)
    =&~
    \rho_{33}
    &\Leftarrow 
    \vcenter{
      \def\w{15mm}
      \def\h{6mm}
      \xymatrix@C=0mm@R=0mm{
        {\scriptstyle q=1} &
        *=<\w,\h>[F]{
          0
        } 
        & 
        *=<\w,\h>[F]{
          \rho_{33}
        } 
        &
        *=<\w,\h>[F]{
          1
        } 
        \\ 
        {\scriptstyle q=0} &
        *=<\w,\h>[F]{
          1
        } 
        &
        *=<\w,\h>[F]{
          \mathemph{
            \rho_{33} 
          }
        } 
        &
        *=<\w,\h>[F]{
          0
        } 
        \\ 
        & {\scriptstyle p=0} & {\scriptstyle p=1} & {\scriptstyle p=2} 
      }}
    \\
    H^1\Big(\Xt, V_2 \otimes V_{1}^\dual\Big)
    =&~
    RG^{\oplus 16}
    &\Leftarrow 
    \vcenter{
      \def\w{15mm}
      \def\h{6mm}
      \xymatrix@C=0mm@R=0mm{
        {\scriptstyle q=1} &
        *=<\w,\h>[F]{
          0
        } 
        & 
        *=<\w,\h>[F]{
          0
        } 
        &
        *=<\w,\h>[F]{
          0
        } 
        \\ 
        {\scriptstyle q=0} &
        *=<\w,\h>[F]{
          0
        } 
        &
        *=<\w,\h>[F]{
          \mathemph{
            RG^{\oplus 16} 
          }
        } 
        &
        *=<\w,\h>[F]{
          RG^{\oplus 4}
        } 
        \\ 
        & {\scriptstyle p=0} & {\scriptstyle p=1} & {\scriptstyle p=2} 
      }}
    \\
    H^2\Big(\Xt, V_2 \otimes V_{1}^\dual\Big)
    =&~
    RG^{\oplus 4}
    &\Leftarrow 
    \vcenter{
      \def\w{15mm}
      \def\h{6mm}
      \xymatrix@C=0mm@R=0mm{
        {\scriptstyle q=1} &
        *=<\w,\h>[F]{
          0
        } 
        & 
        *=<\w,\h>[F]{
          0
        } 
        &
        *=<\w,\h>[F]{
          0
        } 
        \\ 
        {\scriptstyle q=0} &
        *=<\w,\h>[F]{
          0
        } 
        &
        *=<\w,\h>[F]{
          RG^{\oplus 16}
        } 
        &
        *=<\w,\h>[F]{ 
          \mathemph{
            RG^{\oplus 4}
          }
        } 
        \\ 
        & {\scriptstyle p=0} & {\scriptstyle p=1} & {\scriptstyle p=2} 
      }}
  \end{aligned}
  \,,
\end{equation}
where we marked the relevant entry in the corresponding tableau in
\textcolor{red}{\bf bold face}. Hence, the product map, \eref{eq:cup},
simplifies to
\begin{multline}
  \label{eq:cup2}
  \wedge: 
  H^1\Big(\Xt, \pi_{2\ast}\big( V_2 \otimes V_2^\dual \big) \Big)^\z3z3 
  \otimes
  H^1\Big(\Xt, \pi_{2\ast}\big( V_2 \otimes V_1^\dual \big) \Big)^\z3z3  
  \to \\ \to
  H^2\Big(\Xt, \pi_{2\ast}\big( V_2 \otimes V_1^\dual \big)
  \Big)^\z3z3  
  \,.
\end{multline}
These cohomology spaces are, in turn, determined by the Leray spectral
sequence corresponding to the $\beta_2:B_2\to \IP^1$ fibration:
\begin{equation}
  \begin{aligned}
    H^1\Big(\Xt, \pi_{2\ast}\big( V_2 \otimes V_2^\dual \big) \Big)
    =&~
    \rho_{33}
    &&\Leftarrow 
    \vcenter{
      \def\w{15mm}
      \def\h{6mm}
      \xymatrix@C=0mm@R=0mm{
        {\scriptstyle t=1} &
        *=<\w,\h>[F]{
          \mathemph{
            RG^{\oplus 3}
          }
        } 
        & 
        *=<35mm,\h>[F]{
          0
        } 
        \\ 
        {\scriptstyle t=0} &
        *=<\w,\h>[F]{
          1
        } 
        &
        *=<35mm,\h>[F]{
          \mathemph{
            \scriptstyle
            (\chi_1+\chi_1^2)(1+\chi_2+\chi_2^2)
          }
        } 
        \\ 
        & {\scriptstyle s=0} & {\scriptstyle s=1}
      }}
    \\
    H^1\Big(\Xt, \pi_{2\ast}\big( V_2 \otimes V_1^\dual \big) \Big)
    =&~
    RG^{\oplus 16}
    &&\Leftarrow 
    \vcenter{
      \def\w{15mm}
      \def\h{6mm}
      \xymatrix@C=0mm@R=0mm{
        {\scriptstyle t=1} &
        *=<\w,\h>[F]{
          \mathemph{
            RG^{\oplus 16}
          }
        } 
        & 
        *=<\w,\h>[F]{
          RG^{\oplus 4}
        } 
        \\ 
        {\scriptstyle t=0} &
        *=<\w,\h>[F]{
          0
        } 
        &
        *=<\w,\h>[F]{
          0
        } 
        \\ 
        & {\scriptstyle s=0} & {\scriptstyle s=1}
      }}
    \\
    H^2\Big(\Xt, \pi_{2\ast}\big( V_2 \otimes V_1^\dual \big) \Big)
    =&~
    RG^{\oplus 4}
    &&\Leftarrow 
    \vcenter{
      \def\w{15mm}
      \def\h{6mm}
      \xymatrix@C=0mm@R=0mm{
        {\scriptstyle t=1} &
        *=<\w,\h>[F]{
            RG^{\oplus 16}
          }
        & 
        *=<\w,\h>[F]{
          \mathemph{
            RG^{\oplus 4}
          }
        } 
        \\ 
        {\scriptstyle t=0} &
        *=<\w,\h>[F]{
          0
        } 
        &
        *=<\w,\h>[F]{
          0
        } 
        \\ 
        & {\scriptstyle s=0} & {\scriptstyle s=1}
      }}
  \end{aligned}
  \,,
\end{equation}
where we notice that only the $R^1\beta_{2\ast}$ push-down contributes
to the invariant part of the cohomology spaces. Hence, the product map,
\eref{eq:cup2}, simplifies once more to
\begin{multline}
  \label{eq:cup3}
  \wedge: 
  H^0\Big(\Xt, 
  R^1\beta_{2\ast} \pi_{2\ast}\big( V_2 \otimes V_2^\dual \big) \Big)^\z3z3 
  \otimes 
  H^0\Big(\Xt, 
  R^1\beta_{2\ast} \pi_{2\ast}\big( V_2 \otimes V_1^\dual \big) \Big)^\z3z3 
  \to \\ \to 
  H^1\Big(\Xt, 
  R^1\beta_{2\ast} \pi_{2\ast}\big( V_2 \otimes V_1^\dual \big)
  \Big)^\z3z3
  \,.
\end{multline}
But this product is now zero for degree reasons: the product of two
degree $0$ cohomology spaces is again of degree $0$, and not $1$.
Therefore, the product map specified in
eqns.~\eqref{eq:cup3},~\eqref{eq:cup2}, and~\eqref{eq:cup} is the zero
map. That is,
\begin{equation}
  \delta_1=0
  \,,
\end{equation}
as claimed.

\end{appendices}

\bibliographystyle{JHEP} \renewcommand{\refname}{Bibliography}
\addcontentsline{toc}{section}{Bibliography} \bibliography{moduli}

\end{document}